\documentclass[twocolumn,showpacs,preprintnumbers,amsmath,amssymb]{revtex4}

\usepackage{graphicx}
\usepackage{dcolumn}
\usepackage{bm}
\usepackage{color}

\usepackage{epsfig}
\usepackage{amssymb,amsmath}

\pdfoutput=1

\begin{document}

\preprint{APS/123}
\title{\Large Thermodynamics of Blue Phases In Electric Fields}

\author{O. Henrich$^1$, D. Marenduzzo$^1$, K. Stratford$^2$, M. E. Cates$^1$}
\affiliation{
$^1$ SUPA, School of Physics and Astronomy,
The University of Edinburgh, JCMB, King's Buildings, Mayfield Road,
Edinburgh EH9 3JZ, UK \\
EPCC, The University of Edinburgh, JCMB, King's Buildings, Mayfield Road,
Edinburgh EH9 3JZ, UK}

\date{\today}

\begin{abstract}
We present extensive numerical studies to determine the 
phase diagrams of cubic and hexagonal blue phases in an electric field. We 
confirm the earlier prediction that hexagonal phases, both 2 and 3 dimensional,
are stabilized by a field, but we significantly refine the phase boundaries, 
which were previously estimated {\textcolor{black}{by 
means of a semi-analytical approximation}}.
In particular, our simulations show that the blue phase I -- blue phase II
transition at fixed chirality is largely unaffected by electric field, 
as observed experimentally.
\end{abstract}

\maketitle

\section{Introduction}

The blue phases (BPs) of chiral molecules provide spectacular
examples of soft solids, formed via a spontaneously occurring network 
of disclination lines within a cholesteric background~\cite{Wright1989}. All
BPs arise close to the transition between the cholesteric and the
isotropic phase; in at least two cases (BPI and BPII) the disclination network 
forms an ordered (cubic) lattice. (A third, BPIII, is thought to be
 amorphous~\cite{Wright1989}.) 
When an electric field is applied, several new blue phase disclination
lattices have either been observed or theoretically predicted.
Two of these have hexagonal symmetries, and have been named H$_{2D}$ and 
H$_{3D}$ respectively: these have first been predicted 
theoretically~\cite{Hornreich1985,Hornreich1987} and subsequently seen
experimentally~\cite{Pieranski1985}. A further BP appearing in a field, BPX,
has tetragonal symmetry: it was found experimentally in the 1980s~\cite{Pieranski1987} 
and its structure recently possibly revealed by simulations~\cite{Alexander2008}. For a review of experimental results on blue phases, especially 
in electric fields,
the interested reader should consult Refs.~\cite{Kitzerow1991,Crooker1989}.

In the last few years, BPs have emerged as very promising device materials 
with fast (sub-$\mu$s) and tunable color response. This increase in 
technological potential comes from recent remarkable
experiments~\cite{Coles2005,Kikuchi2002,Kikuchi2007}, which managed to stabilize BPs over
a temperature range of 50 K, compared to ranges of about 1 K previously 
\cite{Wright1989}. However, for their potential to be
fully realized, our understanding of BPs needs to become as quantitative as
the one we have for conventional nematic liquid crystals.

In particular, our theoretical knowledge of the phase diagrams of BP-forming
liquid crystals in the presence of an applied field 
is currently quite dated. It relies on the seminal papers 
by Hornreich {\em et al.}~\cite{Hornreich1985,Hornreich1987}, which are 
however based on an approximation (to allow semi-analytical progress): 
the tensorial liquid crystalline order parameter is represented by
a Fourier series, 
{\textcolor{black}{comprising harmonics of relatively high order
but only corresponding to the $m=2$ helicity mode,
which eliminates the dependence on one of the elastic constants
(see Refs.~\cite{Grebel1984,HornreichPRA1990} for details).}} 
These papers have provided a very useful first
theory of blue phases, but it has become clear recently that their
quantitative predictive power is somewhat 
limited~\cite{Grebel1984,Dupuis2005a,Alexander2006,Alexander2008}. 
For instance, within these approximations, the phase diagram at zero electric
field predicts that BPII and BPI appear in the wrong order 
upon varying the molecular chirality~\cite{Grebel1984} (cf. experimental
phase diagrams shown in, e.g., Ref.~\cite{Crooker1989}). 
At the same time, these theories were also unable to account for anomalous 
electrostriction of BPI~\cite{Stark1991,Zelazna1998,Alexander2008}. 
Recent simulations have shown
that these shortfalls are a drawback of the 
{\textcolor{black}{approximations}} employed, and
not of the underlying Landau -- de Gennes mean field free energy~\cite{deGennes}. The latter is the starting point for describing the thermodynamics of BPs, and when handled directly gives good semi-quantitative agreement with the experiments. The simulations of Refs.~\cite{Dupuis2005b,Alexander2006} use a lattice Boltzmann (LB) method which can address the full dynamics of the system including fluid flow. Recent work favors instead a hybrid method~\cite{Henrich2009a,Cates2009} in which an LB code for fluid degrees of freedom is coupled to a conventional finite difference code for the order parameter (this method was recently used in binary fluids as well~\cite{Tiribocchi2009}). In the current work, we address only thermodynamic steady states, in which case it is convenient to switch off the fluid motion and use only diffusive relaxation to find local minima of the free energy. (We can then compare these minima to construct the phase diagram.) As well as being directly comparable with experiments, the resulting phase diagrams form a secure foundation for future hydrodynamic simulations on the same systems using the hybrid LB approach~\cite{Cates2009}.

Our program in this paper is therefore to update the phase diagrams
for blue phases in an electric field by computer simulation of the governing equations derived from the Landau -- de Gennes free energy functional (detailed below). We find that the existing semi-analytical approximations~\cite{Hornreich1985,Lubin1987,HornreichPRA1990} are able to
capture rather well the qualitative physics of the problem. 
In particular, we confirm that an intermediate electric field stabilizes the 
two previously proposed hexagonal phases, H$_{2D}$ and H$_{3D}$.
However, our simulations show that the blue phases, whether cubic
or hexagonal, extend to regions of 
significantly lower chirality than predicted by the
{\textcolor{black}{semi-analytical}} approximation; this in agreement with the trend which
was observed in phase diagrams of BPs without any electric field~\cite{Dupuis2005a,Alexander2006}.
Furthermore, we also resolve which of the cubic
blue phases, BPI or BPII, is stable at any given point in the phase
diagram. Hence we find that the Landau -- de Gennes free energy leads to
a near-horizontal phase boundary in the field-temperature plane, 
in agreement with experimental observations~\cite{Stegemeyer1984}.

It is important to note that, while in this work we focus on a continuum 
description of blue phase via a Landau -- de Gennes free energy,
there also exist simulation studies of individually resolved 
molecules, described as spherocylinders interacting with a coarse grained
potential~\cite{Memmer2000}, which managed to stabilize blue phases. 
This approach is extremely interesting, however the length and time scales
accessible within it are significantly smaller than the ones we can
cover with our approach, and it would be unfeasible to compute a phase
diagram with this method.

This paper is organized as follows. In Section II we discuss the
Landau -- de Gennes theory which leads to the continuum equation of motion
which we numerically solve. In Section III we present the numerical results, 
i.e. the phase diagrams in the chirality-temperature and 
temperature-field, planes, and we quantitatively compare them
with the ones predicted via semi-analytical approximations. Finally, Section
IV contains our discussion and conclusions.

\section{Landau -- de Gennes Theory}

\subsection{Equations of motion and free energy}

The nematodynamic description of liquid crystals, often named after its
inventors as the Erickson-Leslie-Parodi approach, uses a fixed-magnitude (headless) unit vector 
field as order parameter. This ``director field'' represents the average 
orientation of the liquid crystal molecules. 
However, this approach proves to be inappropriate for BPs, as it cannot 
account for disclination lines: on these topological defects, the strength of nematic ordering goes to zero and no average local orientation can be 
defined. A suitable description of BPs is therefore only possible within the framework 
of the Landau -- de Gennes theory.
This employs an order parameter ${\bf Q}$, which is a traceless 
and symmetric second rank tensor; this can describe simultaneously the direction and magnitude of local nematic ordering.
The Q-tensor approach also allows for this ordering to become biaxial; in general the order parameter can be written as:
\begin{equation}
{\bf Q}=q_l(\vec{l}\otimes \vec{l})+q_m(\vec{m}\otimes \vec{m}) 
-\frac{1}{3}(q_l+q_m) {\bf I},
\end{equation}
where $\otimes$ denotes the tensorial product.
The vector fields $\vec{l}$ and $\vec{m}$ are two independent director fields perpendicular to each other. The quantities $q_l$ and $q_m$ are called the scalar order parameters and ${\bf I}$ is the unit tensor. This representation confirms ${\bf Q}$ to be a traceless symmetric second rank tensor which has five independent components. In many systems the smaller of the two scalar order parameters is everywhere very small (if not strictly zero). A director field $\vec{n}$ and a scalar order parameter $q$ can then be 
reintroduced within the {\em uniaxial approximation}:
\begin{equation}\label{uniaxial-approx}
{\bf Q}\approx q(\vec{n}\otimes \vec{n}) -\frac{q}{3} {\bf I}. 
\end{equation}
This now has three independent components to describe both the strength and the director of the uniaxial ordering. However, we have no need for the uniaxial approximation in the numerical approach used in this paper.

The phenomenological Landau -- de Gennes free energy functional
in an external electric field ${\mathbf E}$ is given as:
\begin{eqnarray}\label{fe-functional}
{\cal F}[{\bf Q}]&=&\int d^3r\left\{\frac{A_0}{2}\left(1-\frac{\gamma}{3}\right)
Q^2_{\alpha \beta} -\frac{A_0\gamma}{3}Q_{\alpha \beta} Q_{\beta \gamma}Q_{\gamma \alpha}\right.\nonumber\\
&+&\left.\frac{A_0\gamma}{4} (Q^2_{\alpha \beta})^2  
-\frac{\epsilon_a}{12 \pi} E_\alpha Q_{\alpha \beta} E_\beta \right.\nonumber\\
&+&\left.\frac{K}{2}\left[(\varepsilon_{\alpha \gamma \delta} \partial_{\gamma} Q_{\delta \beta} + 2 q_0 Q_{\alpha \beta})^2
+(\partial_{\alpha}Q_{\alpha\beta})^2\right]\right\} 
 \end{eqnarray}
The thermodynamic equilibrium state is the global minimum of ${\cal F}$ for any given parameters; metastable phases are local minima of ${\cal F}$.
The complete free energy is made up of three different contributions.
The first one (comprising all terms containing $A_0$) is the bulk free energy, which contains terms of the Q-tensor 
up to fourth order.
The scale factor $A_0$ is called the bulk free energy constant, while the 
parameter $\gamma$ plays the role of an effective reciprocal temperature. 
For a nematogen without chirality ($q_0=0$), for $\gamma < 2.7$ the isotropic state gives
the global minimum of the free energy, 
whereas for $\gamma\ge 2.7$ the system has nematic order in equilibrium.
The relative local distortion enters the free energy functional 
on the level of first order gradient terms; throughout this paper we consider only the ``one elastic constant'' approximation~\cite{deGennes}, which is a 
common approach when 
investigating generic liquid crystallizing behavior. $K$ is the resulting single elastic constant; without this approximation, Eq.~\ref{fe-functional} becomes considerably more complicated.
We have chosen a specific representation of the gradient free energy \cite{Wright1989}, which incorporates a gradient-independent part of the bulk free energy to ensure that it is always positive. 
The parameter $q_0=\frac{2 \pi}{p_0}$ determines the intrinsic preferred pitch length $p_0$ of the underlying cholesteric ({\em i.e.}, chiral nematic) liquid crystal.
Thirdly, the coupling to an external electric field $E_\alpha$ is provided by the remaining term in Eq. \ref{fe-functional}, with $\epsilon_a>0$ being the dielectric anisotropy. 
This term is linear in the Q-tensor and quadratic in the electric field (the latter is dictated by symmetry for non-ferroelectric liquid crystals, as considered here).

For the purposes of finding local minima of ${\cal F}$ it is sufficient to ignore fluid flow and momentum conservation, instead taking the tensor order parameter ${\bf Q}$ to obey the following 
(purely relaxational) equation of motion: 
\begin{equation}\label{op-eom}
\frac{\partial {\bf Q}}{\partial t} = \Gamma {\bf H}.
\end{equation}
The molecular field ${\bf H}$ is defined as the functional derivative of the 
Landau -- de Gennes free energy functional Eq. \ref{fe-functional} with respect to
the order parameter, and therefore it vanishes in equilibrium. It is
specifically given by:
\begin{equation}
{\bf H}=-\frac{\delta {\cal F}}{\delta {\bf Q}}+ \frac{\bf I}{3}\; 
Tr\left(\frac{\delta {\cal F}}{\delta {\bf Q}}\right)
\end{equation}
The parameter $\Gamma$ is a rotational diffusion constant.
(Note that in practice Eq.~\ref{op-eom} is supplemented by an additional dynamical update rule for a so-called `redshift' factor, described below.)

It is convenient to render the free energy functional Eq. \ref{fe-functional} 
dimensionless. This gives rise to the following minimal set of parameters on which phase behavior can depend:
\begin{eqnarray}
\tau&=&\frac{27(1-\gamma/3)}{\gamma}\label{tau}\\
\kappa&=&\sqrt{\frac{108\ K\, q_0^2}{A_0\, \gamma}}\label{kappa}\\
e^2&=&\frac{27 \epsilon_a }{32 \pi A_0 \gamma} \;E_\alpha E_\alpha\label{e}.
\end{eqnarray}
Note that the cholesteric pitch parameter $q_0$ is used in the non-dimensionalization of lengths; clearly this works only for nonzero $q_0$ which we from now on assume. (There are no BPs for achiral molecules with $q_0 = 0$.)
The quadratic term of the dimensionless bulk free energy becomes then 
proportional to the reduced temperature $\tau$, whereas the 
magnitude of the gradient free energy term is proportional  to $\kappa$, the chirality. 
Note that the chirality $\kappa$ expresses a ratio between the gradient free 
energy and the bulk free energy terms and is a measure of how much twist the system wants to have.
The parameter $e$ is an effective field strength. 

The free energy functional Eq.~\ref{fe-functional} yields a highly complicated 
free energy landscape. Although the equilibrium state for each $\tau,\kappa,e$ is uniquely defined by 
the functional, it is not generally possible to determine these states by any analytical minimization procedure. 
An exception is the limit of infinite chirality, where the bulk as well as the external 
field free energy terms become negligible compared to the gradient free energy 
and an analytical solution can be found.
While this limit of infinite chirality is hardly relevant for a real-world 
understanding or experiments, the topological character of the equilibrium defect structure in this limit is often unaltered upon reducing chirality -- so long as the unit of length (related to $q_0$) is rescaled appropriately. 
Therefore, these analytical solutions offer a useful starting point for numerical solutions of the relaxational dynamics: one can initialize the runs at finite chirality with the solution that is already known for the infinite chirality limit. The chosen dynamics will relax this structure towards a local free energy minimum; {\textcolor{black}{we then compare the minimized free energy with those found from competing starting structures to find the best possible local minimum among the ones corresponding to the topologies which we have considered}}. 

\subsection{Initial conditions}

In this work we consider four such structures, alongside the isotropic phase, the (para-)nematic phase, and the standard uniaxial configuration for the cholesteric state: BPI, BPII, H$_{2D}$ and H$_{3D}$ (in this last case we consider two possible starting conditions, labeled H$_{3D}^a$ and H$_{3D}^b$). In each case the relevant infinite-chirality expression was taken as initial conditions. For BPI we used:
\begin{eqnarray}\label{bp1-init}
Q_{xx}&\simeq& -2 \cos(q_0' y)\sin(q_0' z)+\nonumber\\
&& \sin{(q_0' x)} \cos{(q_0' z)}+ \cos{(q_0' x)} \sin{(q_0' y)}\nonumber\\
Q_{xy}&\simeq& \sqrt{2} \cos{(q_0' y)} \cos{(q_0' z)}+\nonumber\\
&& \sqrt{2} \sin{(q_0' x)} \sin{(q_0' z)}- \sin{(q_0' x)} \cos{(q_0' y)} \nonumber\\
Q_{xz}&\simeq&\sqrt{2}\cos(q_0' x) \cos(q_0' y)+\nonumber\\
&& \sqrt{2} \sin(q_0' z) \sin(q_0' y)- \cos(q_0' x) \sin(q_0' z)\nonumber\\ 
Q_{yy}&\simeq&-2 \sin(q_0' x)\cos(q_0' z)+\nonumber\\
&& \sin(q_0' y)\cos(q_0' x)+  \cos(q_0' y)\sin(q_0' z) \nonumber\\
 Q_{yz}&\simeq& \sqrt{2} \cos(q_0' z) \cos(q_0' x)+\nonumber\\
&&  \sqrt{2}\sin(q_0' y) \sin(q_0' x)- \sin(q_0' y)\cos(q_0' z),
\end{eqnarray}
where $q_0'=\sqrt{2} q_0$. For BPII we used:
\begin{eqnarray}\label{bp2-init}
Q_{xx}&\simeq& \cos(2 q_0 z)-\cos(2 q_0 y);\;\;\; Q_{xy}\simeq \sin(2 q_0 z)\nonumber\\
Q_{xz}&\simeq& \sin(2 q_0 y); \;\; Q_{yy}\simeq \cos(2 q_0 x)-\cos(2 q_0 z)\nonumber\\
Q_{yz}&\simeq& \sin(2 q_0 x).
\end{eqnarray}
For the hexagonal two-dimensional BP, abbreviated as H$_{2D}$, we used: 
\begin{eqnarray}\label{h2d-init}
Q_{xx}&\simeq& -\frac{3}{2} \cos(q_0 x) \cos(\sqrt{3} q_0 y)\nonumber\\
Q_{xy}&\simeq& -\frac{\sqrt{3} }{2} \sin(q_0 x) \sin(\sqrt{3}q_0 y)\nonumber\\
Q_{xz}&\simeq&\sqrt{3} \cos(q_0 x) \sin(\sqrt{3}q_0 y)\nonumber\\
Q_{yy}&\simeq& - \cos(2 q_0 x)-\frac{1}{2} \cos(q_0 x) \cos(\sqrt{3}q_0  y)\nonumber\\
Q_{yz}&\simeq&-\sin(2 q_0 x)-\sin(q_0 x) \cos(\sqrt{3}q_0  y).
\end{eqnarray}
Finally, there are two possible hexagonal three-dimensional BPs identified in
Ref.~\cite{HornreichPRA1990}, which were found to be stable in
the case of positive dielectric constant which we consider here. 
For the first one, abbreviated as H$_{3D}^a$, we used:
\begin{eqnarray}\label{h3d-init}
Q_{xx}&\simeq& -\frac{3}{2}\cos(q_0 x) \cos(\sqrt{3} q_0 y)+\frac{1}{4}\cos(q_0 z)\nonumber\\
Q_{xy}&\simeq&-\frac{\sqrt{3} }{2} \sin(q_0 x) \sin(\sqrt{3} q_0 y)+\frac{1}{4} \sin(q_0 z)\nonumber\\
Q_{xz}&\simeq& \sqrt{3} \cos(q_0 x) \sin(\sqrt{3} q_0 y)\nonumber\\
Q_{yy}&\simeq&-\cos(2 q_0 x)- \frac{1}{2} \cos(q_0 x) \cos(\sqrt{3} q_0 y)- \frac{1}{4}\cos(q_0 z)\nonumber\\
Q_{yz}&\simeq&-\sin(2 q_0 x)-\sin(q_0 x) \cos(\sqrt{3} q_0 y).
\end{eqnarray}
The starting condition for the second three-dimensional 
hexagonal phase, abbreviated as H$_{3D}^b$, is the same as 
the one used for H$_{3D}^a$, with the difference that all terms 
depending on either $x$ or $y$ (or both) have the opposite sign.

In order to calculate the phase diagrams in the three-dimensional parameter 
space spanned by $\tau, \kappa$ and $e^2$, we follow the evolution of the 
free energy for the different configurations towards their (in general 
metastable) equilibria by solving Eq. \ref{op-eom}.
Once the ${\bf Q}$ tensor reached a stationary state, we compare the 
free energy densities of the various structures and identify the lowest one as giving the true equilibrium. 
As already mentioned, this ignores the possibility of still lower values being achieved by phases that we have not considered, of which the most relevant are probably BPX (tetragonal) and BPIII (mentioned further below). In interpreting the phase diagrams shown below, the reader should bear in mind that their validity is thus restricted to the eighth phases (isotropic, nematic, cholesteric, BPI, BPII, H$_{2D}$, and the two H$_{3D}$ phases) detailed above. We have restricted to these phases in line with the previous literature~\cite{Hornreich1985,HornreichPRA1990}: introduction of other topologies may therefore lead to changes in this phase diagram. 
 
An additional limitation to our work is as follows. We have already stated that the pitch parameter $\kappa$ (via $q_0$) sets the preferred length scale for BP structures. However, the dimensionless number that relates the BP unit cell to $q_0$ is not known {\em a priori} but must be found as part of the free energy minimization. In numerical work, however, one can only simulate an integer number of unit cells within a periodic simulation domain. The procedure is then to fix the periodic domain but rescale the size of the physical volume that it represents by a so-called `redshift' factor. One then finds the redshift that gives the lowest free energy as part of the dynamical minimization. For a completely general disclination lattice one requires an independent rescaling of each spatial dimension. For simplicity we here assume a single redshift factor to govern all three dimensions ({\em i.e.}, we assume a `cubic redshift'). This is in line with previous literature practice ~\cite{Lubin1987,Hornreich1985,HornreichPRA1990,HornreichMCLC1990},
and it is the best we can currently do: an anisotropic redshift would effectively require
a separate simulation for each aspect ratio chosen, and this would not be feasible 
with our current computational power. The assumption of a single redshift is exact for BPI and BPII only so long as these remain strictly cubic structures. In the presence of a field, both should acquire a tetragonal distortion which we cannot allow for; however, this effect is thought to be small~\cite{Hornreich1985}. The two H$_{3D}$ structures are handled exactly only if the chosen aspect ratio of the unit cell box (which, for a cubic finite difference grid as used here, must be rational) coincides with the true one. Again, the error with our chosen value (26/15, see below) is expected to be minor.

It is useful to briefly sketch how the algorithm to determine the redshift works. The
method we use was originally proposed in Ref.~\cite{Alexander2006}.
The free energy functional Eq. \ref{fe-functional} comprises terms up to quadratic order in 
gradients of the Q-tensor: upon a rescaling of the unit cell dimension, $L\to L/r$,
it is easy to see that the gradient term are rescaled by a factor of $r$ per derivative. 
Therefore if we rescale $L\to L/r$, the new free energy becomes:
\begin{equation}
f({\bf Q})= r^2 A(\partial {\bf Q}^2)+ r B(\partial {\bf Q}) +C.
\end{equation}
The scaling factor $r^*$ which minimizes the functional is then simply
\begin{equation}
r^*=-\frac{B}{2 A}.
\end{equation} 
This observation provides a simple recipe to determine the redshift at every time step. 

\subsection{Numerical Aspects}\label{numaspects}

Here we briefly present the computational details of our 
work. The size of the simulation boxes was $L_x=L_y=L_z=32$ lattice sites for the 
cubic BPs BPI and BPII and for the cholesteric phase.
To accommodate the hexagonal BPs we chose boxes of $L_x=52, L_y=L_z=30$ lattice 
sites for H$_{2D}$ (although a cubic box would do in this case) and $L_x=L_z=52, L_y=30$ lattice sites for the two H$_{3D}$ structures.
These values were chosen as the closest approximation to
the ratio $1:\sqrt{3}$ which admits a `perfect' hexagonal lattice such as the HCP structure of hard spheres. (As mentioned above, computational limitations did not allow us to explore different values of this aspect ratio parameter, which might be somewhat different for the true structure, even at $e = 0$.) 
The helical pitch was set to 16 lattice sites for the cubic BPs and for the 
cholesteric phase, and to 15 lattice sites for the hexagonal BPs.
For BPI and BPII we generally simulated a box containing eight unit cells, to check for any (large scale) field-induced reconstruction;
for the hexagonal BPs we used a single unit cell.
As initial configurations we used a simple uniaxial helix for the cholesteric 
phase, together with the infinite chirality solutions 
Eqs.~\ref{bp1-init}-\ref{h3d-init}. 
The equation of motion of the order parameter Eq.~\ref{op-eom} was solved 
by using a finite difference scheme \cite{Marenduzzo2007} and a rotational 
diffusion constant $\Gamma=0.3$ in simulation units.
At each time step, the value of the redshift, equivalent of the 
optimal scaling factor $r^*$, was computed. This requires little 
extra calculation as the quantities $A$ and $B$ (see
Eqs.~13-14) are needed for the free energy. Once we know the
value of $r^*$, the unit cell needs to be ``rescaled''.
Instead of changing the actual simulation cell to be simulated, which
would be inefficient and inaccurate, we rescale the elastic constants and
$q_0$ by $1/r^*$ and $r^*$ respectively.
Typical runs to reach equilibration required 25,000 time steps.
At the end of the run we compared the free energy densities in order to 
determine the equilibrium phase as the one with the lowest free energy density.
To determine the onset of the nematic phase a visual check of the 
director field was performed.

\section{Results}

\subsection{Phase Diagrams}

In this Section we present the phase diagrams obtained for cholesteric blue phases in the presence of
an electric field.
To validate our simulations, we first checked the field-free case, which was published in Ref.~\cite{Alexander2006} 
using a code similar to ours (see also Ref.~\cite{Dupuis2005a} for a version
without the variable redshift).


The left panel curves in Fig. \ref{fig1} were found to agree well with 
previous LB work~\cite{Dupuis2005a,Alexander2006} except for minor deviations at the BPI-BPII boundary which could be explained by a slight difference in accuracy used.
A few general remarks are in order to relate this phase diagrams to experimental ones, as well as to previous analytical predictions.
{\textcolor{black}
{At odds with the phase diagrams seen experimentally, such as the ones shown in \cite{Crooker1989}, another BP, known as BPIII or the ``Blue Fog'' is missing.
In experiments, the BPII region is entirely enclosed between BPI and BPIII, while in Fig. \ref{fig1} it emerges as a region which is open towards 
higher chiralities.
BPIII is believed to be an amorphous BP, which is stable at higher chiralities and is thermodynamically distinct from the others~\cite{Kutnjak1995,Koistinen1995,Crooker1989}.
Although its structure is still a subject of discussion and has not been fully understood yet, 
there is some evidence that it might be closer to BPII than to the isotropic phase~\cite{Crooker1989}. 
For theoretical attempts to understand the
structure of BPIII see e.g. Refs.~\cite{Hornreich1986,Rochsar1986,Englert1998}.}
 In principle our numerical approach {\em does} allow us to study large systems in regions in parameter space for which BPIII is expected, 
so that this limitation can be removed, at the cost of much more expensive simulations. (These must be many times larger than any shown here, so as to contain many cells of the aperiodic structure without strong finite size effects.)  We accordingly defer our investigation of BPIII to a future publication \cite{Henrich2009b}. 

It is interesting to compare our approach to an older realization of the Landau -- de Gennes theory \cite{Grebel1984,Hornreich1985,Hornreich1987}.
Within this approach the existence of field-stabilized hexagonal blue phases was first predicted. As mentioned in the introduction,
the major drawback within this older framework was that, due to computational limitations at 
that time, the Q-tensor had to be expanded in a Fourier-series 
{\textcolor{black}{comprising several harmonics but only the $m=2$ helicity
mode. This simplifications leads to the result that the last elastic term in
Eq. (\ref{fe-functional}) does not contribute to the phase diagram, 
which we believe is at the root of the quantitative discrepancy with our 
simulations.}}
As a consequence the {\textcolor{black}{semi-analytical}} theory found the incorrect sequence of blue phases in the phase diagram.
These earlier results are displayed in the right panel of Fig. \ref{fig1}.
Remarkably, our approach shows instead the correct sequence (cholesteric / BPI / BPII) beneath the boundary to the isotropic state.
Furthermore, in the earlier approximate approach  BPII appears adjacent to the cholesteric phase and BPI occurs at considerably lower temperatures
than in the full numerical phase diagram reported here and in Refs.~\cite{Dupuis2005a,Alexander2006}.
{\textcolor{black}{Another important aspect is that the $O_5$ structure, 
which is predicted theoretically (see e.g. Ref.~\cite{Grebel1984}) but
not observed experimentally so far, is relegated to higher chiralities
than the ones considered in our phase diagrams.}}
For a long time these issues cast some doubt on the appropriateness of the Landau -- de Gennes theory for describing blue phases, but it is now clear  that this
functional is actually in good qualitative agreement with experiments, and that the inconsistencies with observations were
essentially a drawback of the approximations previously employed. 



We turn now to the role of applied fields, $e\neq 0$, which forms the main focus of the current work.
Figures \ref{fig3}-\ref{fig4} show phase diagrams in chirality-temperature plane for increasing field strengths $e=0.2$ and $e=0.3$. 
We can identify three main effects of the external field on the phase boundaries.
First of all, in the region where the isotropic phase appeared in the field-free case, the system becomes oriented under the influence of the
external field, forming a paranematic phase (N). This applies as well for the low-chirality part of the ordered region, where a nematic phase now appears in place of the cholesteric phase.
For increasing field strengths the low-chirality nematic-cholesteric phase boundary and the cholesteric phase itself both move towards the 
right hand side of the phase diagram, {\em i.e.}, to higher chiralities.
Secondly, increasing the value of the electric field leaves the BPI-BPII phase boundary almost unchanged, 
an interesting feature which has also been observed in experiments \cite{Stegemeyer1984}. The extent of the cholesteric phase is more or less 
retained. The BPI region gets therefore overtaken by the cholesteric phase and its phase boundaries move towards lower temperatures.
Finally, very close to the BP-nematic boundary a pocket of hexagonal blue phases opens up.
The right panel in Fig. \ref{fig3} display where Hornreich and Shtrikman 
\cite{Hornreich1985} predicted these stable hexagonal BPs at the same 
field strength.
Our results again suggest that their theoretical approach works well in terms of the general location of these phases, but quantitatively it 
suffers from the limitations of the approximations used.
Our more accurate numerics show that while a transition from the H$_{3D}$ phases to H$_{2D}$ can arise under increasing field strength 
(see Fig.~\ref{fig5} below), the series of transitions most often observed is H$_{3D}^b$ $\to$ H$_{3D}^a$ $\to$ H$_{2D}$.
We further predict that the boundary between H$_{3D}^b$ and BPII, the closest among the cubic BPs,
should be straight both in the $(\kappa,\tau)$ plane.
Finally, we confirmed the observation from earlier work that H$_{2D}$ is closer to the nematic phase with respect to the other hexagonal phases:
this is perhaps expected as H$_{2D}$ has a symmetry plane perpendicular to the field direction and its character is therefore closer to that
of a nematic than that of the other BPs. 
{\textcolor{black}{As the stability region of the $O_5$ structure
is outside the chirality range we considered at zero field, 
we have not pursued its study further at non-zero field (we
expect an electric field to destabilise it, as happens with the other 
cubic phases).}}



Another way to visualize the phase behavior, especially to determine the experimentally relevant 
critical field strength for the switching into the nematic state, is to perform cuts through the $\tau, \kappa, e$ parameter space for selected chiralities $\kappa$.
Fig. \ref{fig5} and \ref{fig6} show phase diagrams in field-temperature plane for $\kappa=1.0$ and $\kappa=1.5$ respectively. 
A comparison with the {\textcolor{black}{semi-analytical}} 
calculations of Ref. \cite{Grebel1984} shows that their phase boundaries are roughly in the same place as ours, except that a BPI phase is entirely missing 
in the relevant parameter region.
The near horizontal BPI- BPII boundary seen experimentally in \cite{Stegemeyer1984} is very well reproduced in our findings.

 
\subsection{Visualization of the BPs}

One of the most common ways to characterize the structure of liquid crystals is to visualize their local order, {\em e.g.,} via their director field.
Within the Q-tensor theory a director can be defined according to Eq. \ref{uniaxial-approx} as the normalized length vector which belongs to the largest eigenvalue of the Q-tensor. 
However, as the preferred configuration in BPs is an arrangement of double twist cylinders involving a highly complex network of disclinations, it is almost impossible to 
visualize their structure in more than two dimensions in this way.
Perhaps the clearest alternative is then to show the defect structure itself, which has become the usual way to portray BP structures.
In our simulations, this can easily be achieved by setting a threshold for the scalar order parameter $q$ (defined as the maximum eigenvalue of the tensor order parameter) and imaging the resultant isosurface, which divides `defect' (low $q$) from `non-defect' (high $q$) zones. Although rather simple, this method works remarkably well and we use it here due to its simplicity of implementation.
The following figures \ref{fig7}-\ref{fig10} show therefore isosurfaces of the scalar order parameter $q$.


Fig. \ref{fig7} represents a unit cell of the hexagonal two-dimensional BP 
H$_{2D}$ in the field-free case $e=0$ at $\tau=0.75, \kappa=1.5$. 
The isosurface correspond to a value $q=0.11$. 
This is a metastable configuration, because the H$_{2D}$ phase is only found to be the equilibrium phase for fields $e\approx0.3$.
Clearly recognizable is the hexagonal arrangement of the disclination lines. 
These are oriented along the $z$-direction, which is the direction of the applied external field.
(With a field on, neither of the hexagonal BPs could be maintained with disclinations perpendicular to the field.) 
Note that a unit cell of the hexagonal structures carries a full turn of the director when one passes through it along the $x$- or $y$-direction, in contrast with a half turn in the cases of the cubic blue phases (this can be seen, e.g., 
from checking the periodicity of the tensorial order parameter in Eqs. 9-12).


If the external field is switched on, as shown in Fig. \ref{fig8}, the regions with low order become slightly larger and the disclination tubes facet.
The magnitude of the order parameter at its maxima is found to be larger than in the field-free case.


The situation is more complicated for the hexagonal three-dimensional BPs, H$_{3D}^a$ and H$_{3D}^b$. 
Fig. \ref{fig9} shows a metastable example of H$_{3D}^a$ 
at $\tau=0.6$, $\kappa=1.5$ and $e=0$; one can readily recognize a
hexagonal arrangement if viewing along the $z$ direction, i.e. perpendicular
to the xy plane.
However, when viewed from other directions 
the structure looks quite intricate. 
The general appearance, but not the topology, of the structure slightly depends on the value of the threshold of the 
scalar order parameter $q$; we have chosen a threshold of $q=0.15$.


Fig. \ref{fig10} shows the same phase with external field. The defect regions 
along the field direction merge and increase in their extent.
Similarly, in the $xy$ symmetry plane  
the domains with higher order around the 
center of the voids increase in size.

Fig. \ref{fig11} and \ref{fig12} instead show the other 3D hexagonal phase,
H$_{3D}^b$, both without and with a field. Again the former configuration
is metastable, while the latter is stable. With respect to H$_{3D}^a$,
the disclination lines making up the H$_{3D}^b$ network do not merge
at any point (at least for the values of the threshold, chosen, 
again $q=0.15$). These lines are twisted and arranged onto parallel
planes, the orientation of neighboring planes being about $60^o$ 
consistently with the overall hexagonal symmetry. When a field is
applied and the H$_{3D}^b$ phase is stable, the disclinations twist up
(this is more evident when viewed in a direction perpendicular to
the $xy$ plane -- compare the top panels of Fig. \ref{fig11} and \ref{fig12}).

The result that the BPI-BPII phase boundary in Fig. \ref{fig5} and \ref{fig6} 
remains almost constant with increasing field strength (before the system
enters first the cholesteric and then the nematic phase) 
has its counterpart in the near constancy of the
disclination network of these structures as the field is varied.
 Indeed the cubic BPs appear to be much more stable than the hexagonal BPs 
and are hardly deformed by the influence of the external field.
The top picture in Fig. \ref{fig13} gives the isosurface $q=0.22$ of the order 
parameter in BPI, whereas the picture at the bottom shows the isosurface 
$q=0.15$ in BPII.
The cyan surfaces depict the field-free case, while the magenta parts show results for the field strength $e=0.3$.
It is clear that the disclination network changes only slightly, and the 
isosurfaces are only slightly displaced in direction of the external field. (Recall however that our single-redshift approach excludes any anisotropic distortion of these structures that could arise at high field strengths.) 
 

\section{Conclusions}

We have presented extensive numerical simulations to investigate
the phase diagram of cholesteric blue phases in an electric field.
We have considered the case of a positive dielectric anisotropy, and we 
have for computational simplicity taken into account only isotropic 
deformations of the unit cell for each structure. While this leaves out by 
necessity phenomena like the ``anomalous electrostriction'' of 
BPI~\cite{Stark1991,Zelazna1998,Alexander2008},
this is a good starting point, particularly for comparisons with the earlier literature which
often employed this approximation. 

We confirm in their main qualitative aspects the predictions of Refs.~\cite{Hornreich1985,Hornreich1987}, which were found by a semi-analytical approximation 
{\textcolor{black}{which led to the phase diagram being independent of
one of the elastic constants}}.
 These works predicted that intermediate values of the
electric field stabilizes two new hexagonal blue phases, one
with 2-dimensional and another one with 3-dimensional symmetry. With
respect to previous literature, we were able to refine the phase diagrams
in two important ways. First, we find that the blue phase regions extends to significantly lower
values of chirality than previously estimated.
Second, we have established which of the cubic blue phases, BPI or BPII, 
is stable for a fixed value of electric field, chirality and temperature.
At a given chirality value, we found that the boundary between BPI and BPII 
does not appreciably depend on temperature, in agreement with 
experimental results. Finally we have seen that the disclination lattices of the hexagonal
blue phases modify quite strongly as the field is increased,
in contrast to what happens for the cubic blue phases which, up until close to the 
transition to the nematic or cholesteric state, are not greatly affected in structure by applied fields.

To improve further the accuracy of our phase diagrams, while remaining within the general framework of the Landau -- de Gennes approach, it would be necessary not only to relax the `cubic redshift' assumption, but also to go beyond the one elastic constant approximation. This would allow further insight into the adequacy of the Landau -- de Gennes theory in quantitatively accounting for the experimental phase behavior. Meanwhile, our work lays a useful foundation for future computer simulations of blue phases in
electric fields, addressing for instance the switching dynamics of blue phase devices. To study dynamics, rather than the equilibrium phase behavior addressed here, one must however use the full momentum-conserving equations of motion. To this end, our finite difference code has already been married to a lattice Boltzmann approach~\cite{Cates2009,Henrich2009a} creating a hybrid code with which we plan to address such issues in the near future.

\begin{acknowledgments}
We acknowledge support by EPSRC grants EP/E045316 and EP/E030173, and
useful discussions with G. P. Alexander and E. Orlandini.
M.E.C. holds a Royal Society Research Professorship.
\end{acknowledgments}

\bibliography{bp_electric_fields}

\newpage

\onecolumngrid

\begin{figure}[h]
\includegraphics[width=0.49\textwidth]{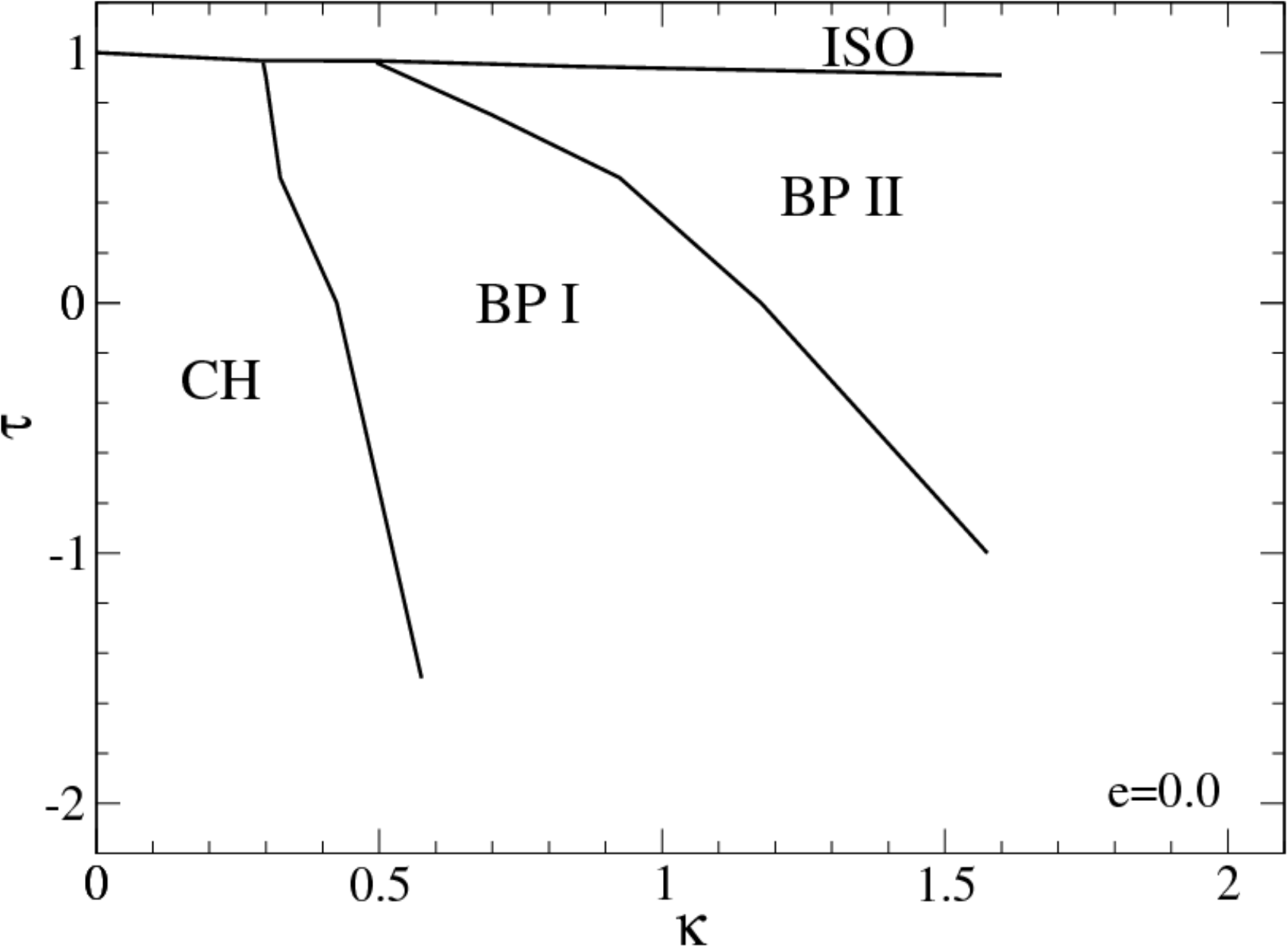}
\includegraphics[width=0.49\textwidth]{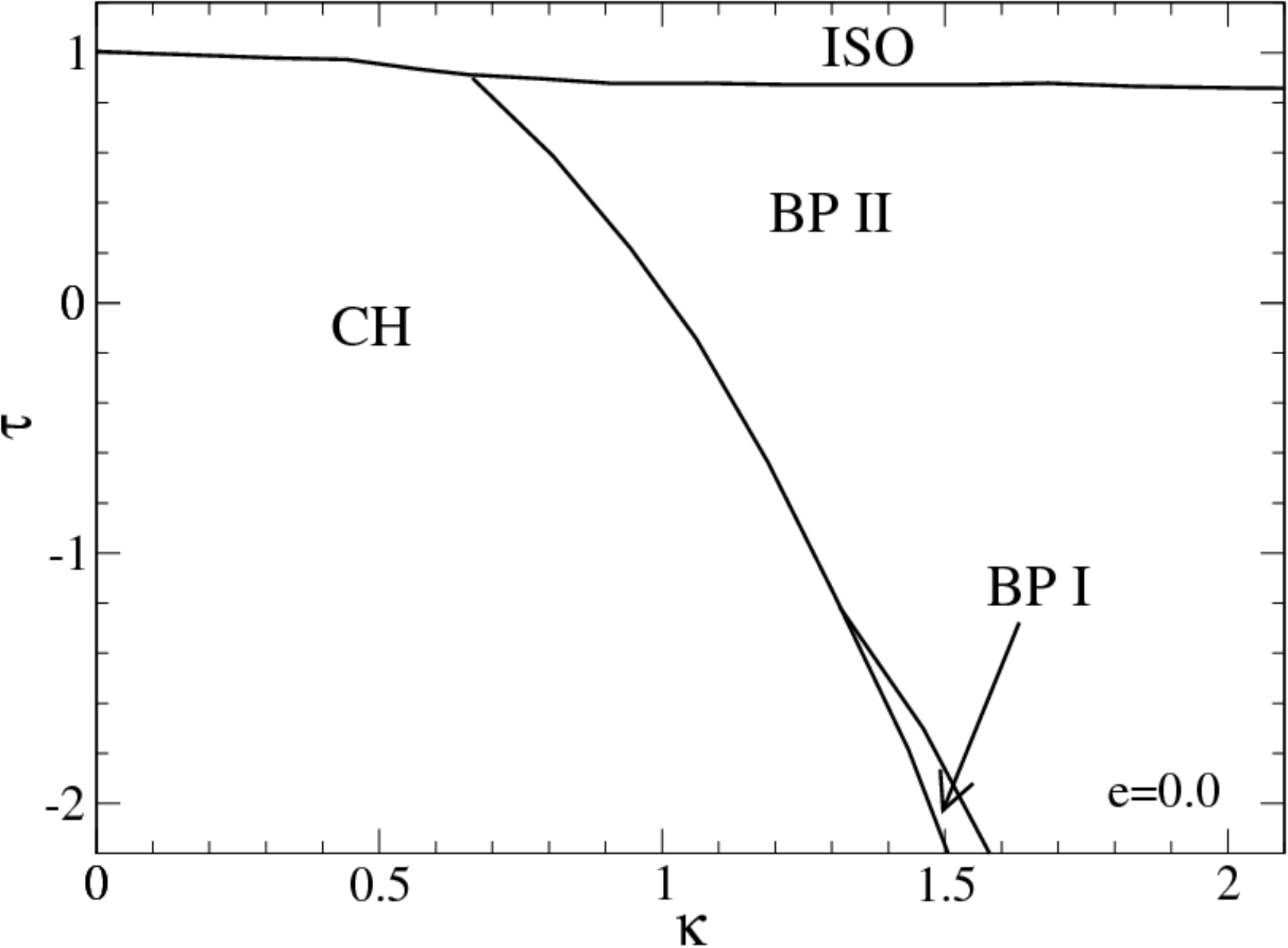}
\caption{Phase diagrams in chirality-temperature plain for $e=0$,
as found within our approach (left) and in Ref.~\cite{Hornreich1985} (right).
{\textcolor{black}{Typically, redshift values for BPI were smaller than
for BPII, e.g. at $\tau=0$, for $\kappa=0.55$ and $\kappa=1.1$ 
BPI was found to be stable, with a redshift $r$ equal to,
respectively, $r=0.7$ and $r=0.76$; while for $\kappa=1.4$ BPII was stable
and $r=0.88$.}} }
\label{fig1}
\end{figure}

\begin{figure}[h] 
\includegraphics[width=0.49\textwidth]{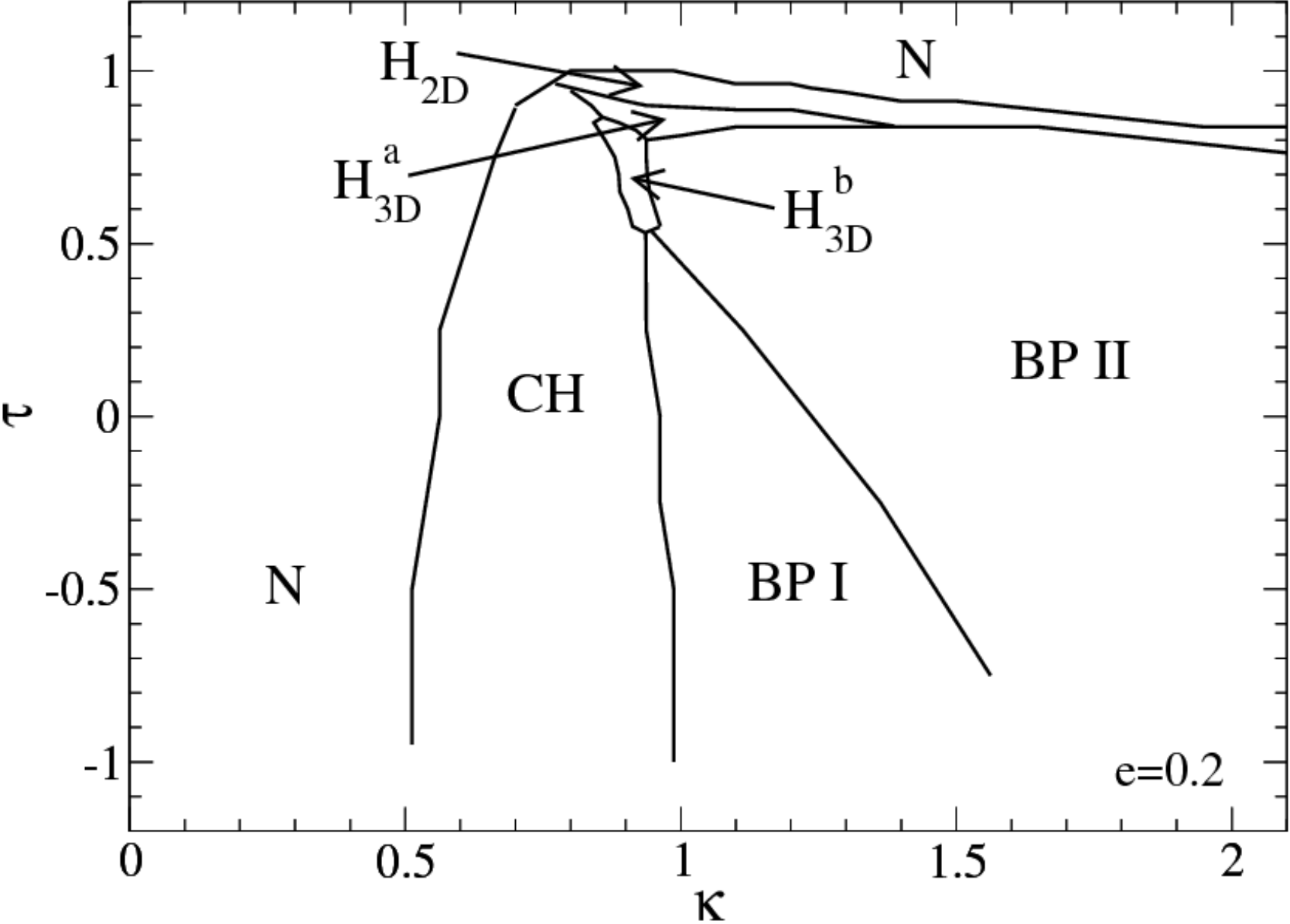}
\includegraphics[width=0.49\textwidth]{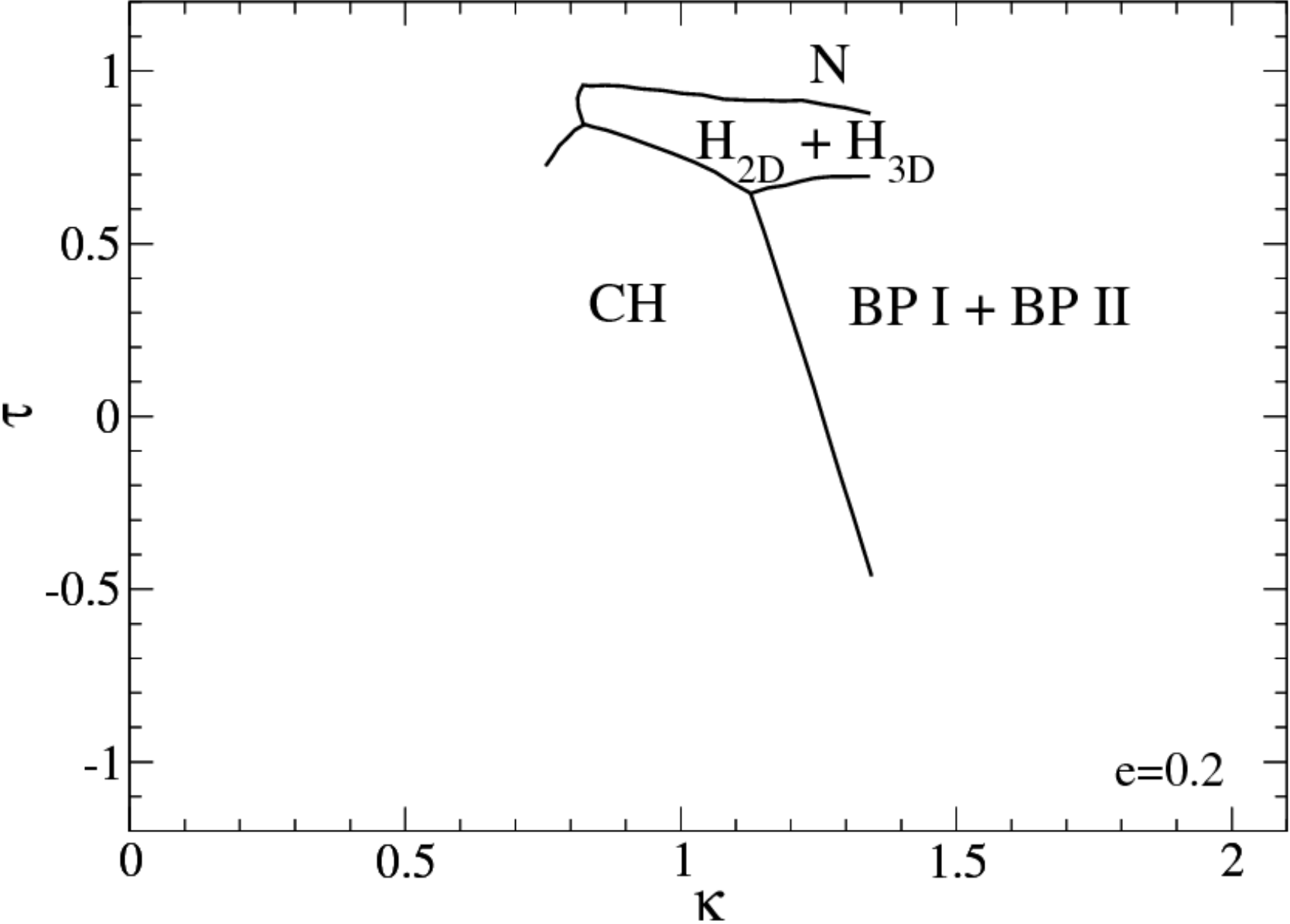}
\caption{Phase diagrams in chirality-temperature plain for $e=0.2$,
as found with our approach (left) and in Ref.~\cite{Hornreich1985} (right).
{\textcolor{black}{Redshift values were typically smaller for BPI than
for BPII, and smaller for cubic blue phases than for hexagonal ones: 
e.g. for BPI 
at $\tau=0$, and $\kappa=1.1$, $r=0.82$;
for BPII at $\tau=0$ and $\kappa=1.4$, $r=0.89$;
for H$_{2D}$ at $\tau=0.95$ and $\kappa=0.9$, $r=0.96$;
for  H$_{3D}^a$ at $\tau=0.9$ and $\kappa=0.9$, $r=0.935$;
for  H$_{3D}^b$ at $\tau=0.8$ and $\kappa=0.9$, $r=0.91$. 
Note that for smaller values of $e$ we reproduce electrostriction
and the redshift decreases albeit very slightly, in agreement with
previous theoretical and numerical 
literature~\cite{Lubin1987,Alexander2008}.}}}
\label{fig3}
\end{figure}

\begin{figure}[h]
\includegraphics[width=0.5\textwidth]{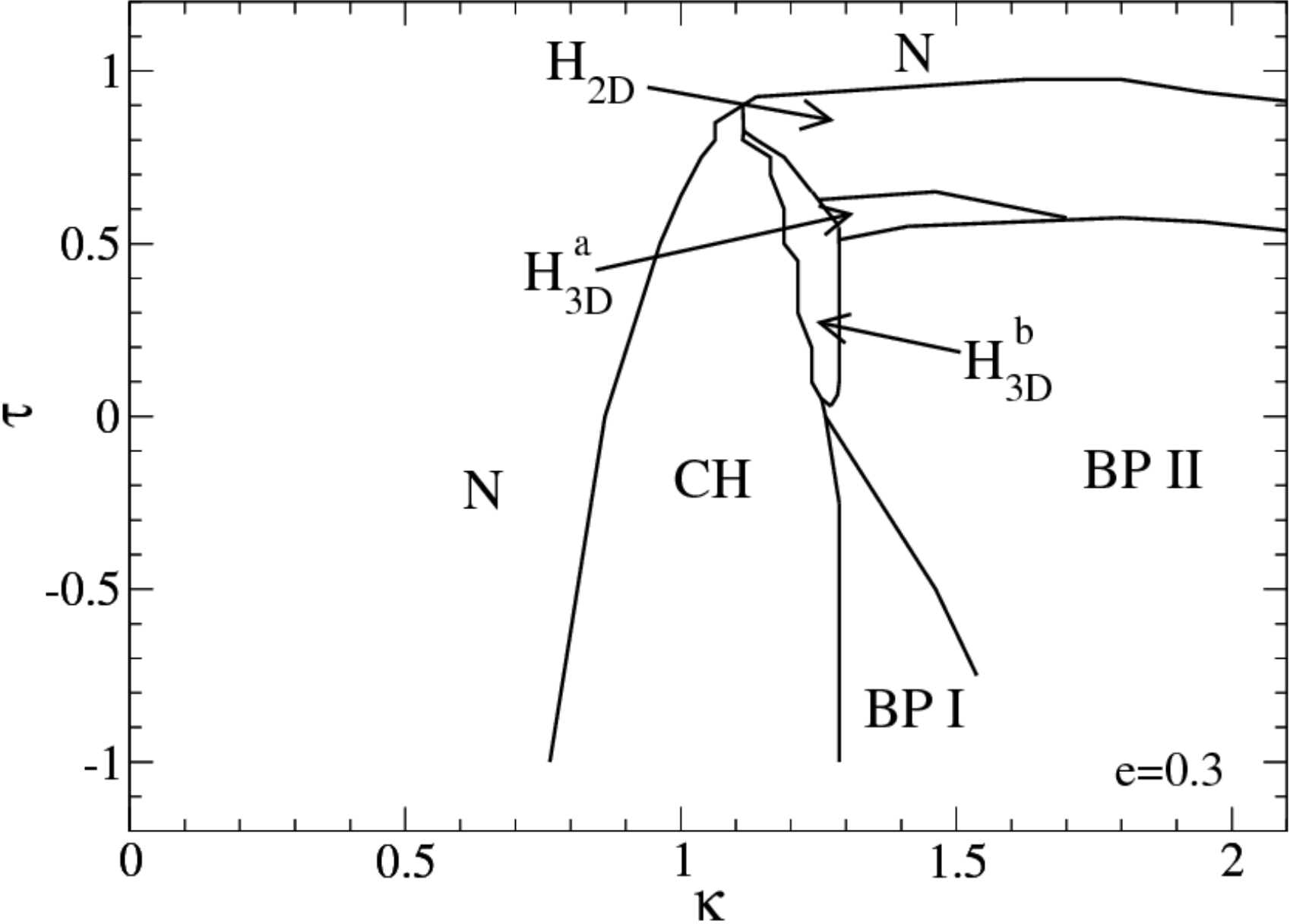}
\caption{Phase diagram in chirality-temperature plain for $e=0.3$.}
\label{fig4}
\end{figure}

\begin{figure}[h]
\includegraphics[width=0.49\textwidth]{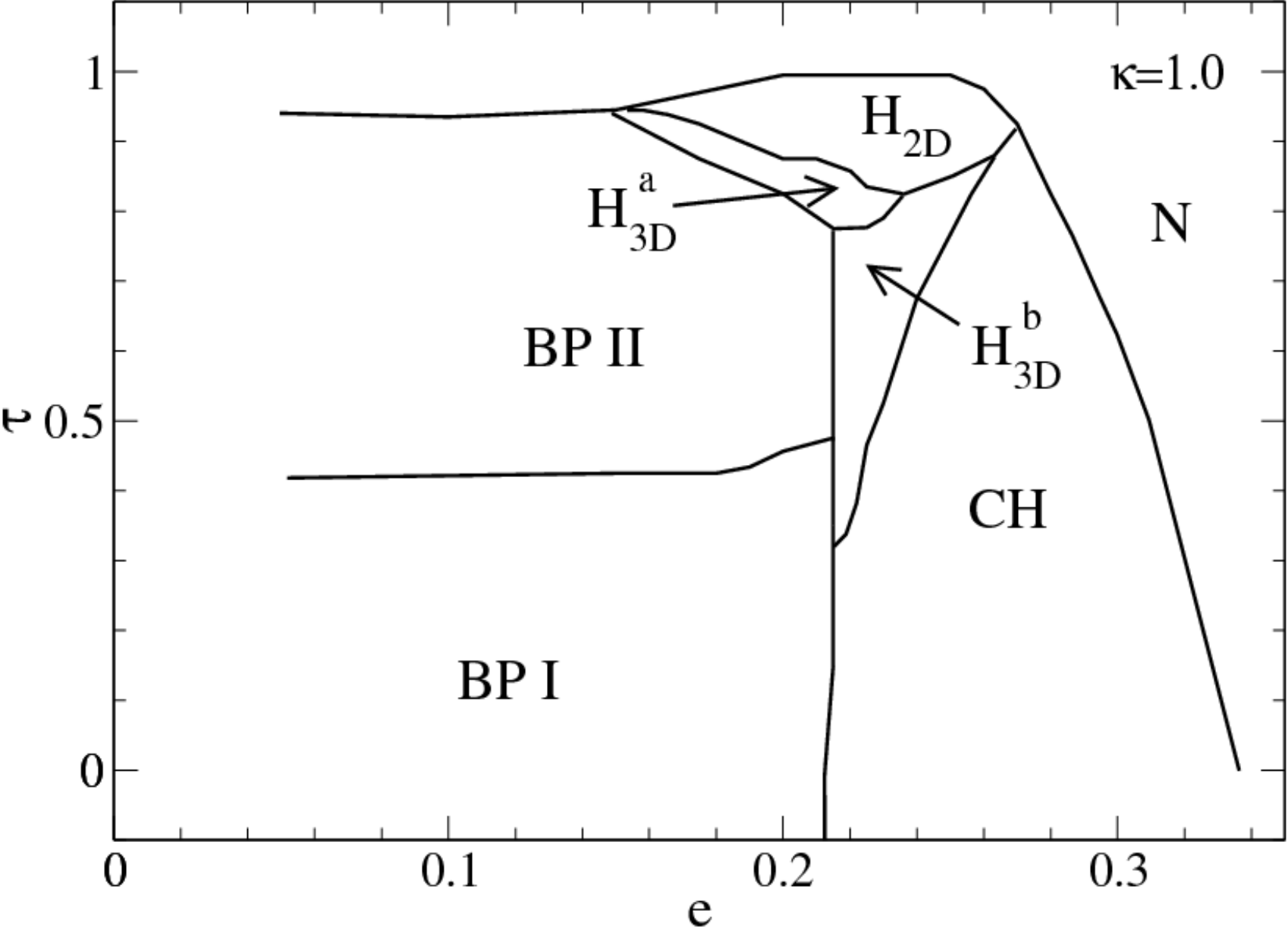}
\includegraphics[width=0.49\textwidth]{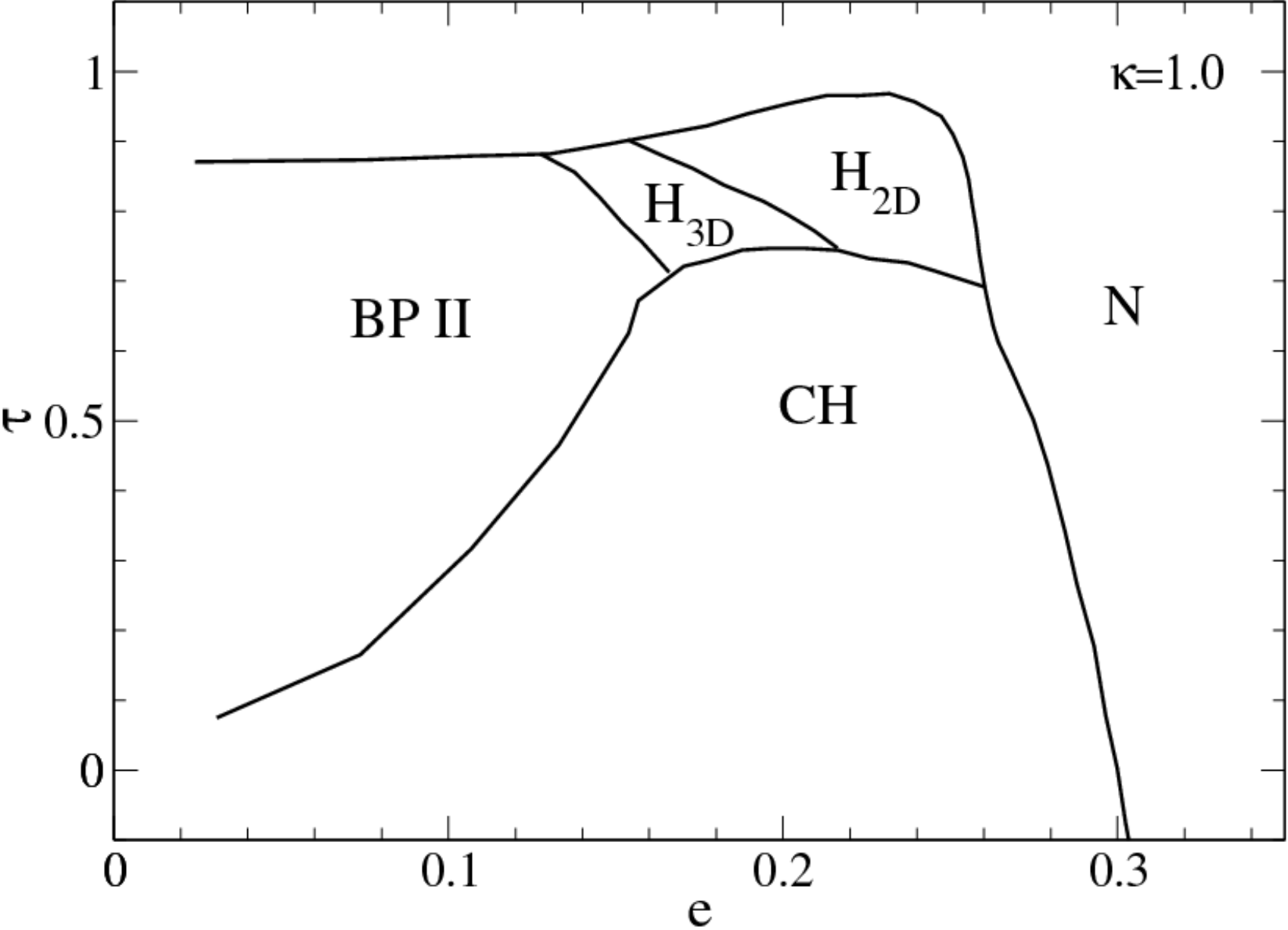}
\caption{Phase diagrams in field strength-temperature plain for $\kappa=1.0$
as found with our approach (left) and in Ref.~\cite{Hornreich1985} (right). 
Note that change in scale for $\tau$.}
\label{fig5}
\end{figure}

\begin{figure}[h]
\includegraphics[width=0.49\textwidth]{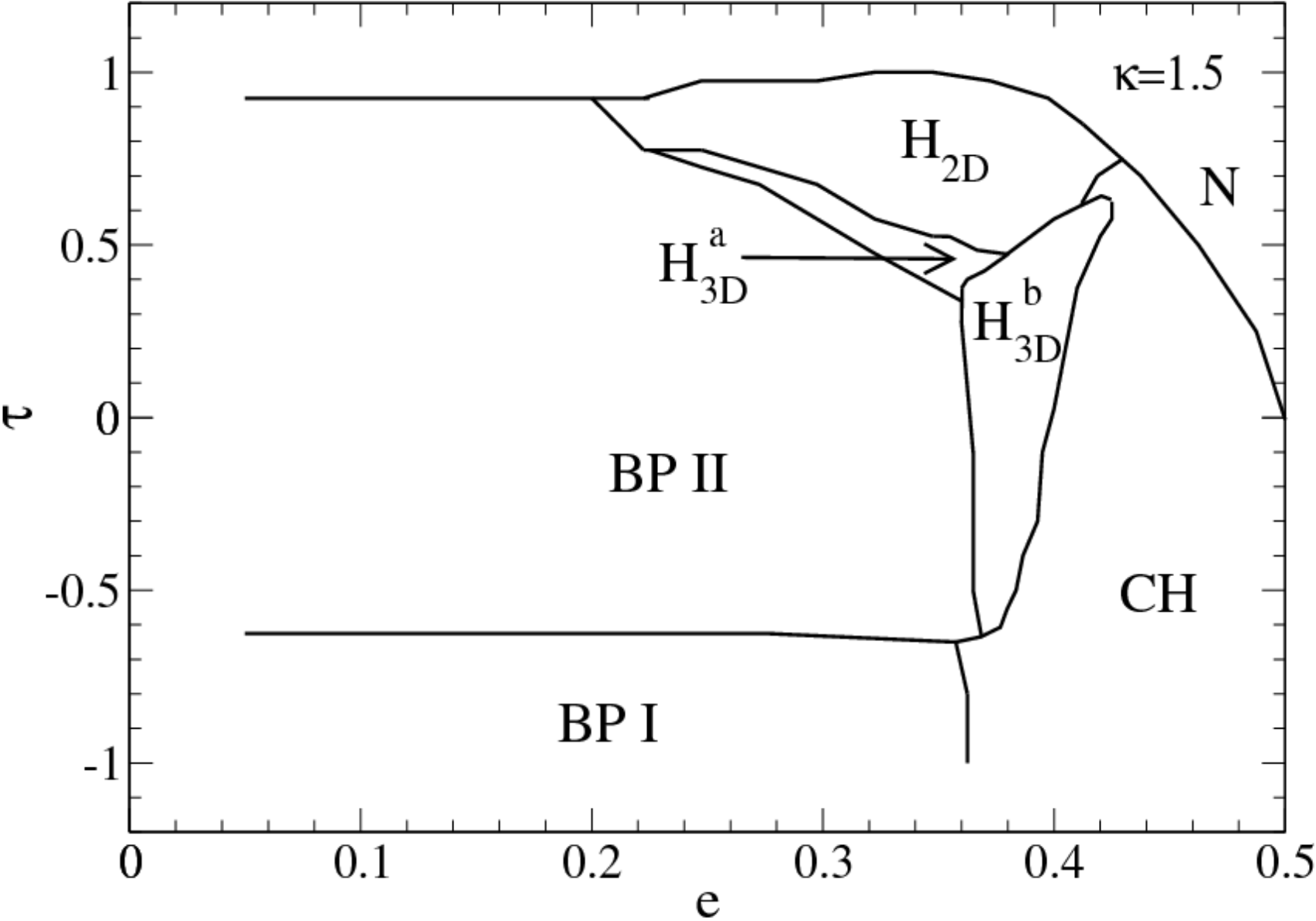}
\includegraphics[width=0.49\textwidth]{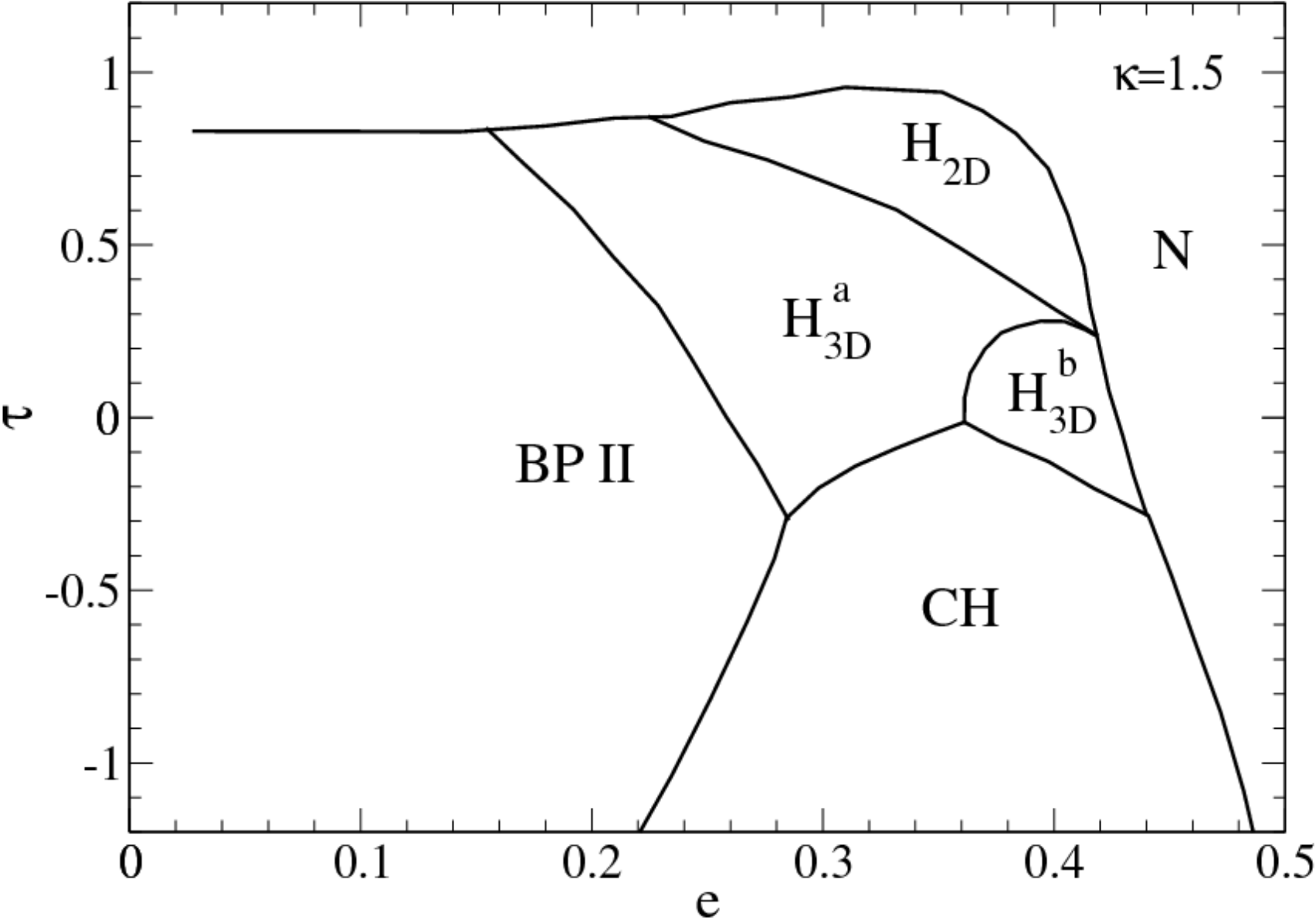}
\caption{Phase diagrams in field strength-temperature plain for $\kappa=1.5$
as found with our approach (left) and in Ref.~\cite{Hornreich1985} (right).}
\label{fig6}
\end{figure}

\twocolumngrid

\begin{figure}[h]
\includegraphics[width=0.5\textwidth]{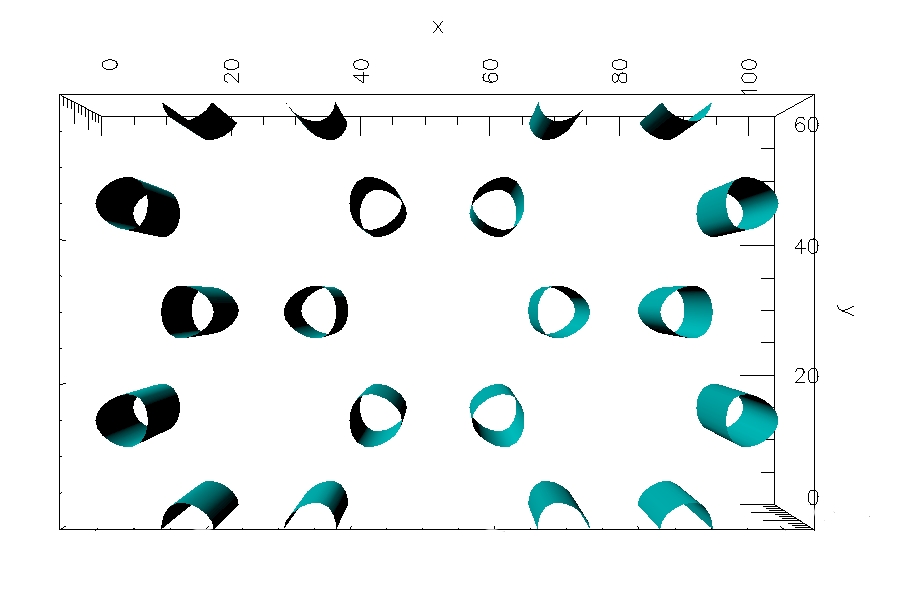}
\includegraphics[width=0.45\textwidth]{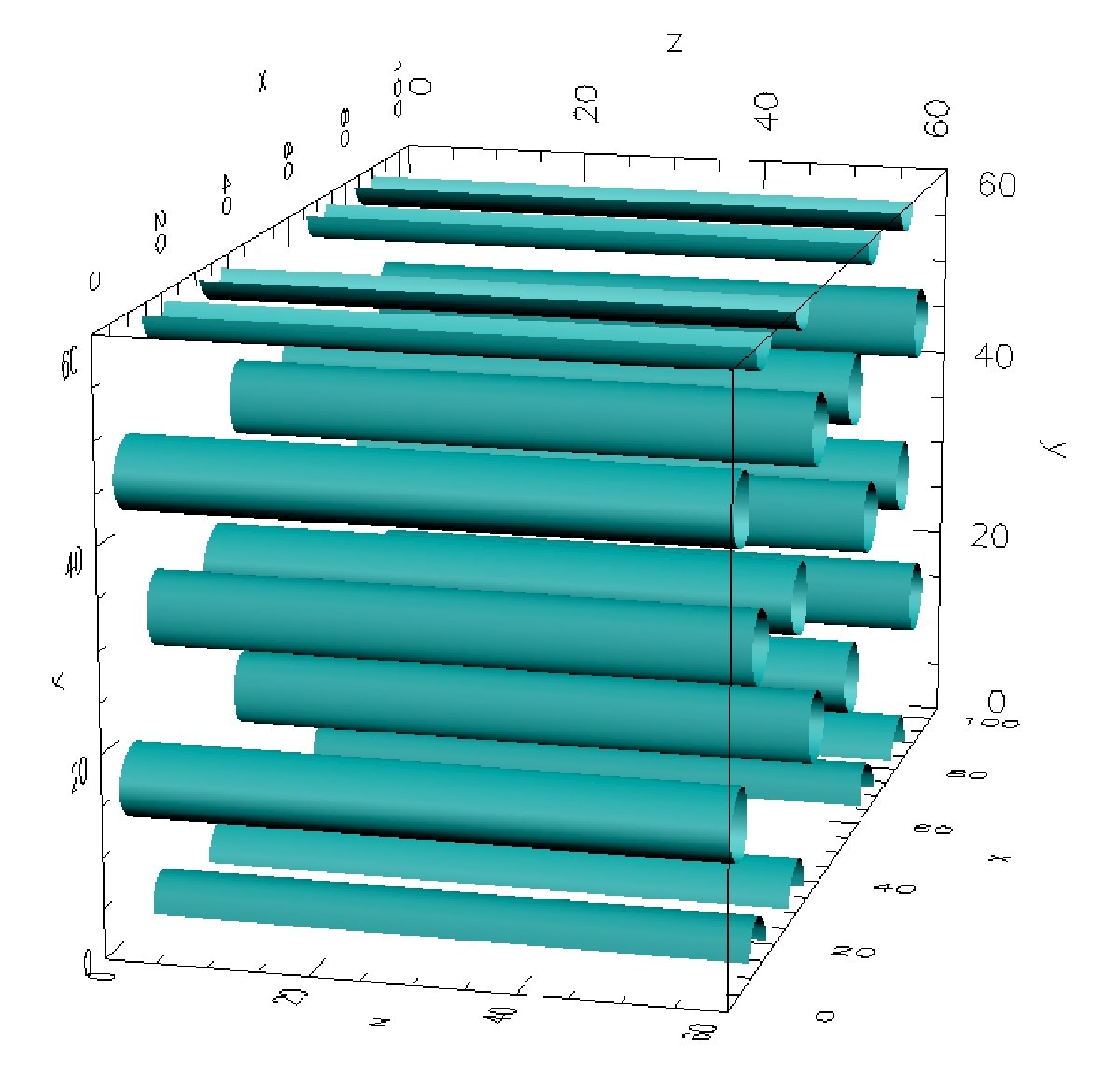}
\caption{Hexagonal-2D BP in equilibrium state at $\tau=0.75, \kappa=1.5, e=0$. Each tube contains a disclination line.}
\label{fig7}
\end{figure}

\begin{figure}[h]
\includegraphics[width=0.5\textwidth]{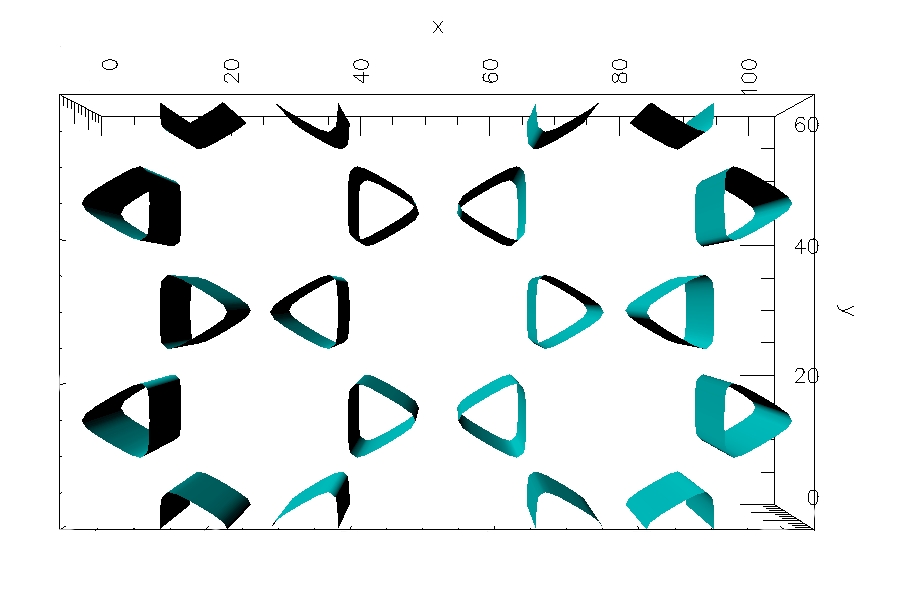}
\includegraphics[width=0.45\textwidth]{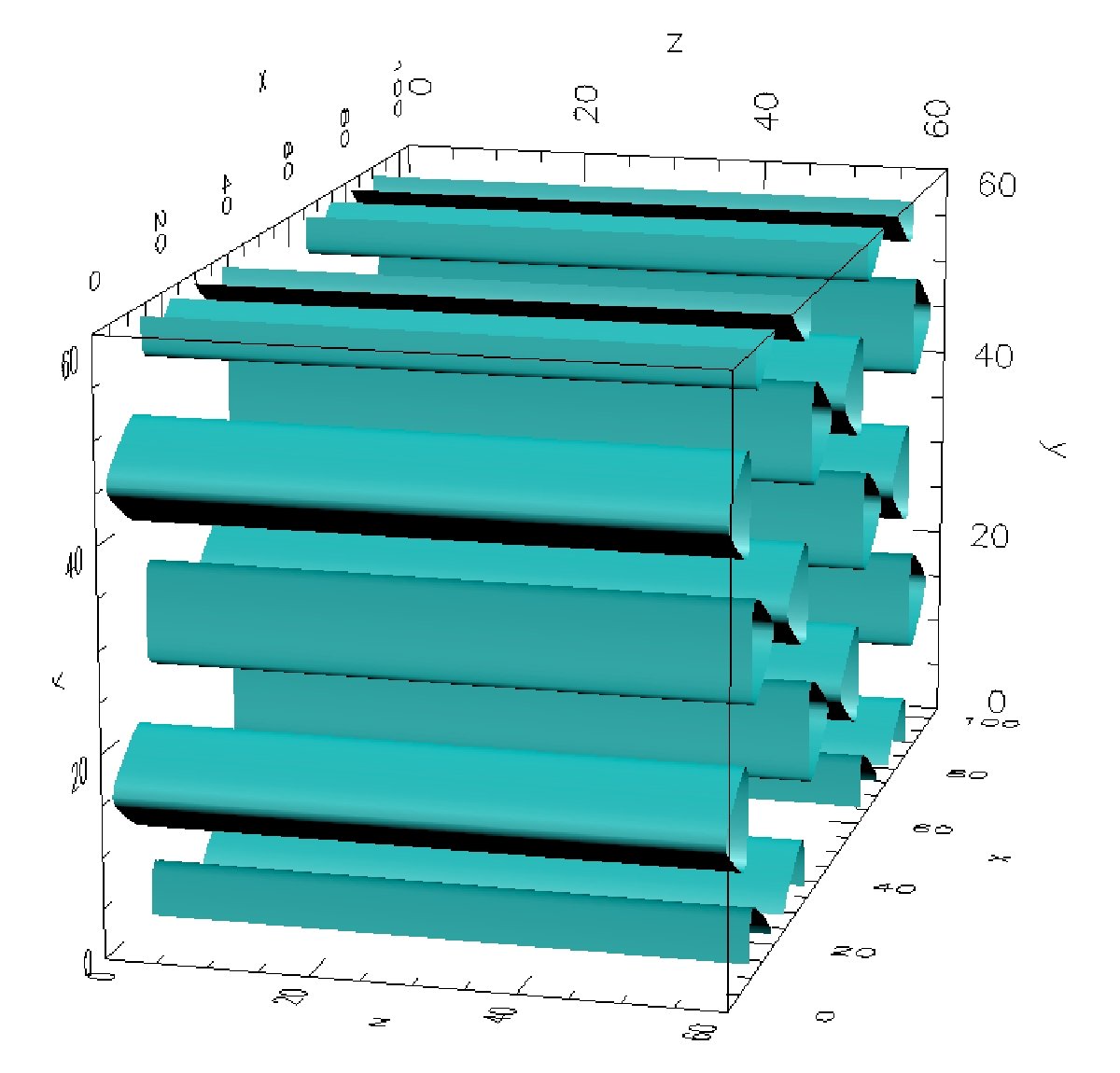}
\caption{Hexagonal-2D BP in equilibrium state at $\tau=0.75, \kappa=1.5, e=0.3$.}
\label{fig8}
\end{figure}

\begin{figure}[h]
\includegraphics[width=0.5\textwidth]{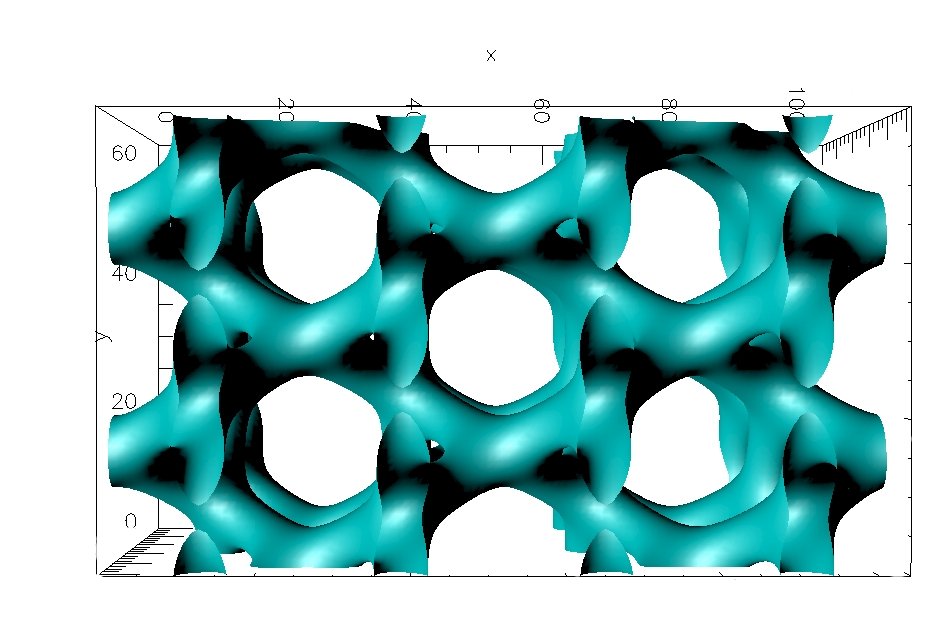}
\includegraphics[width=0.5\textwidth]{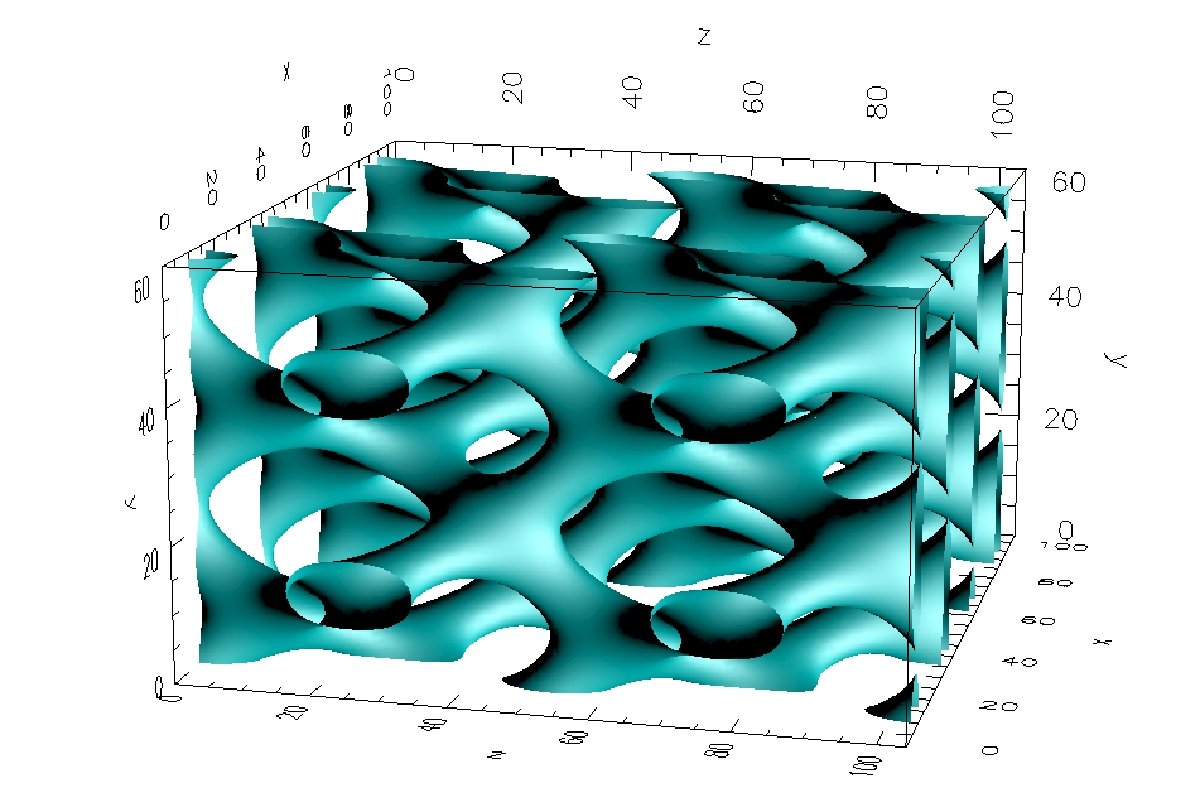}
\caption{Hexagonal-3Da BP in metastable state at $\tau=0.6, \kappa=1.5, e=0$.}
\label{fig9}
\end{figure}

\begin{figure}[h]
\includegraphics[width=0.5\textwidth]{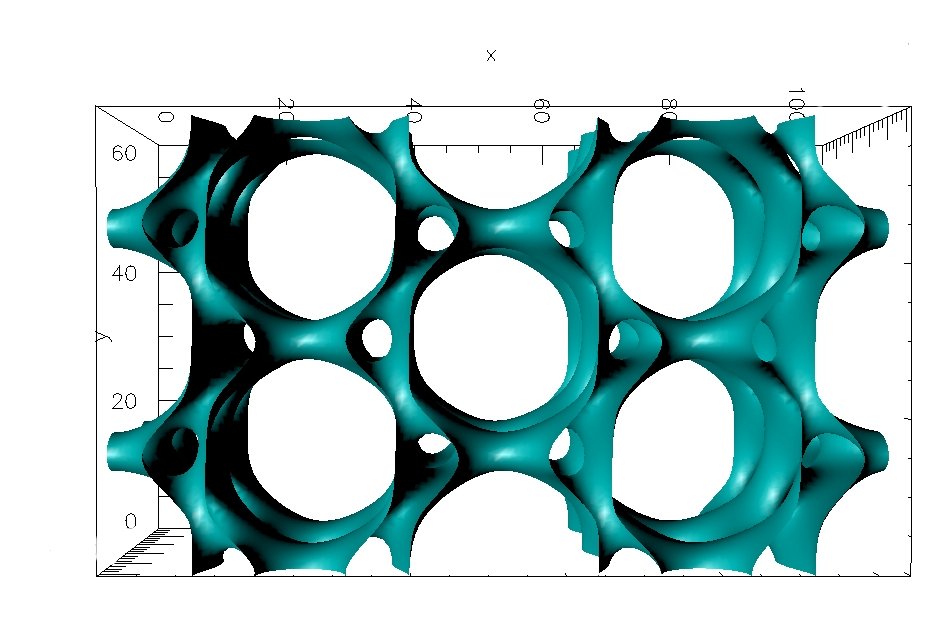}
\includegraphics[width=0.5\textwidth]{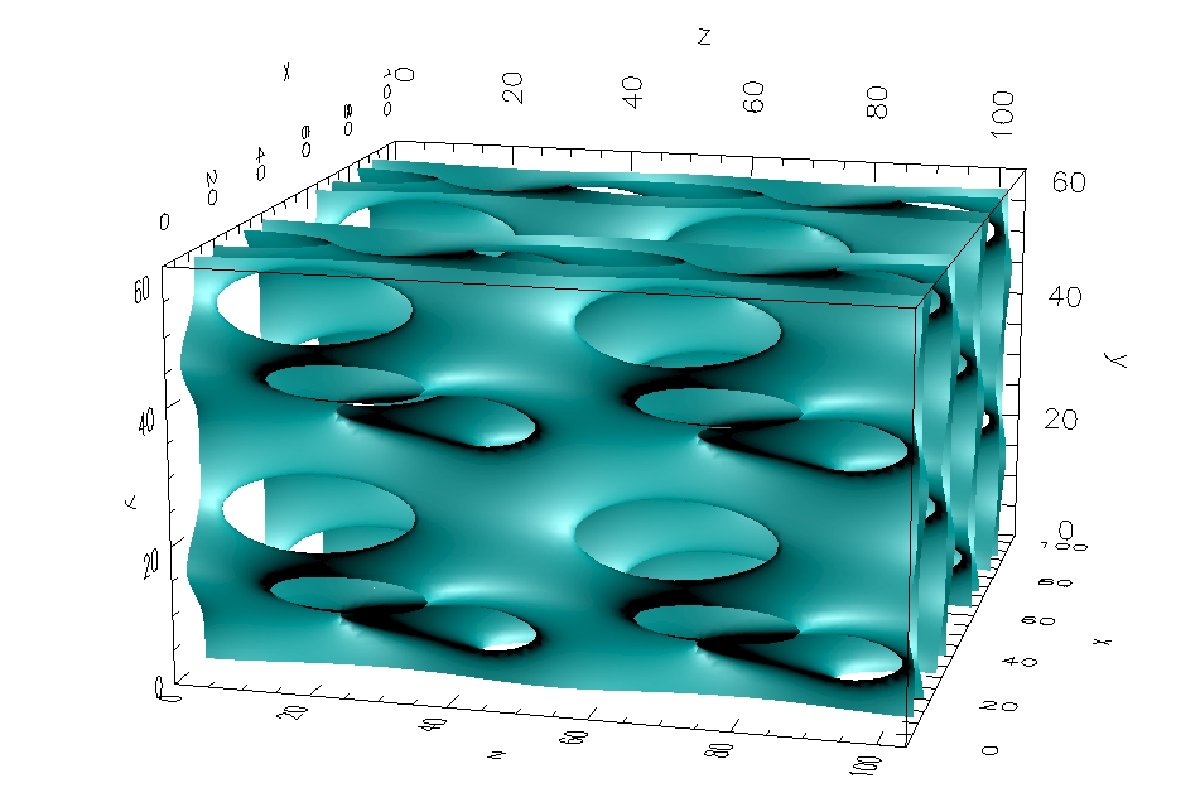}
\caption{Hexagonal-3Da BP in equilibrium state at $\tau=0.5, \kappa=1.25, e=0.3$.}
\label{fig10}
\end{figure}

\begin{figure}[h]
\includegraphics[width=0.5\textwidth]{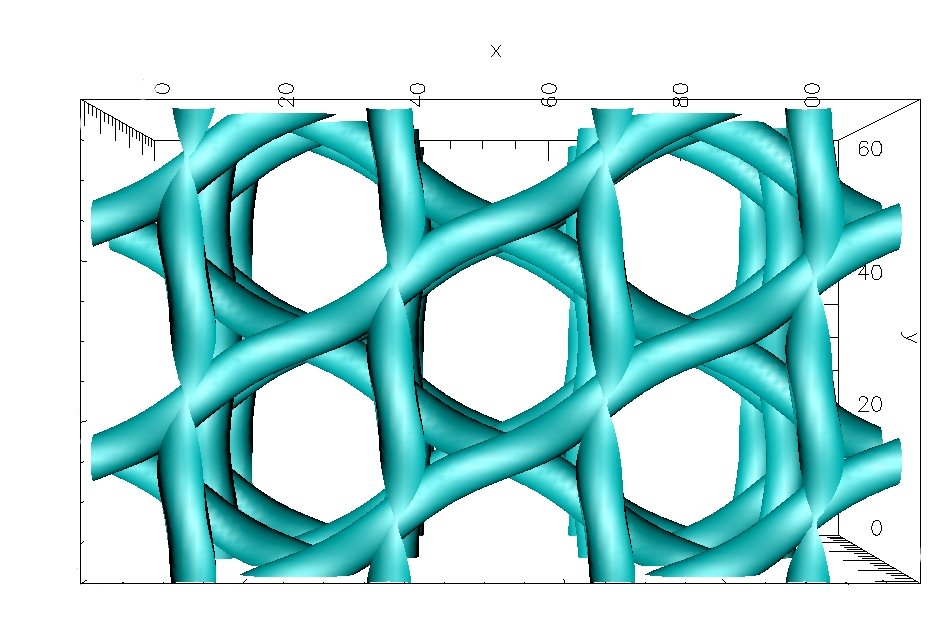}
\includegraphics[width=0.5\textwidth]{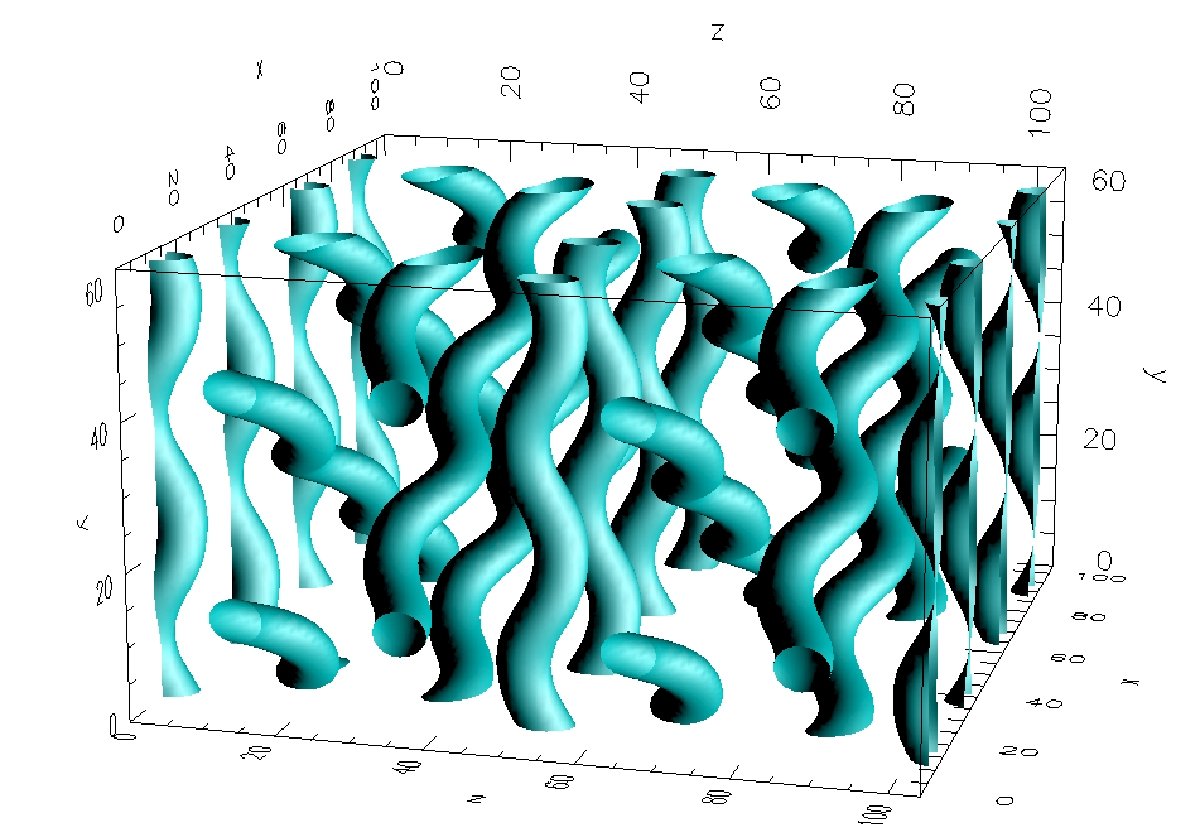}
\caption{Hexagonal-3Db BP in metastable state at $\tau=0.5, \kappa=1.25, e=0$.}
\label{fig11}
\end{figure}

\begin{figure}[h]
\includegraphics[width=0.5\textwidth]{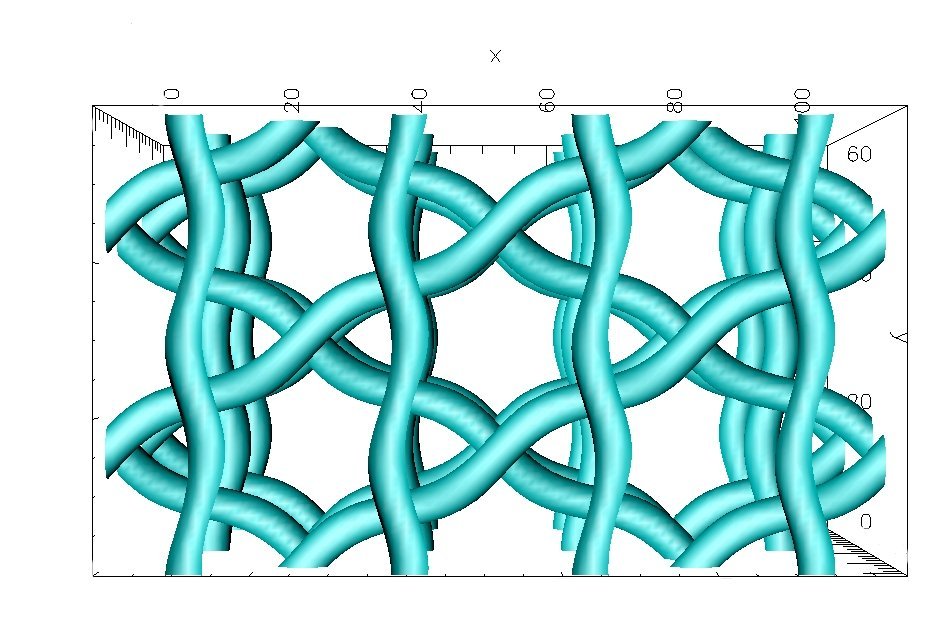}
\includegraphics[width=0.5\textwidth]{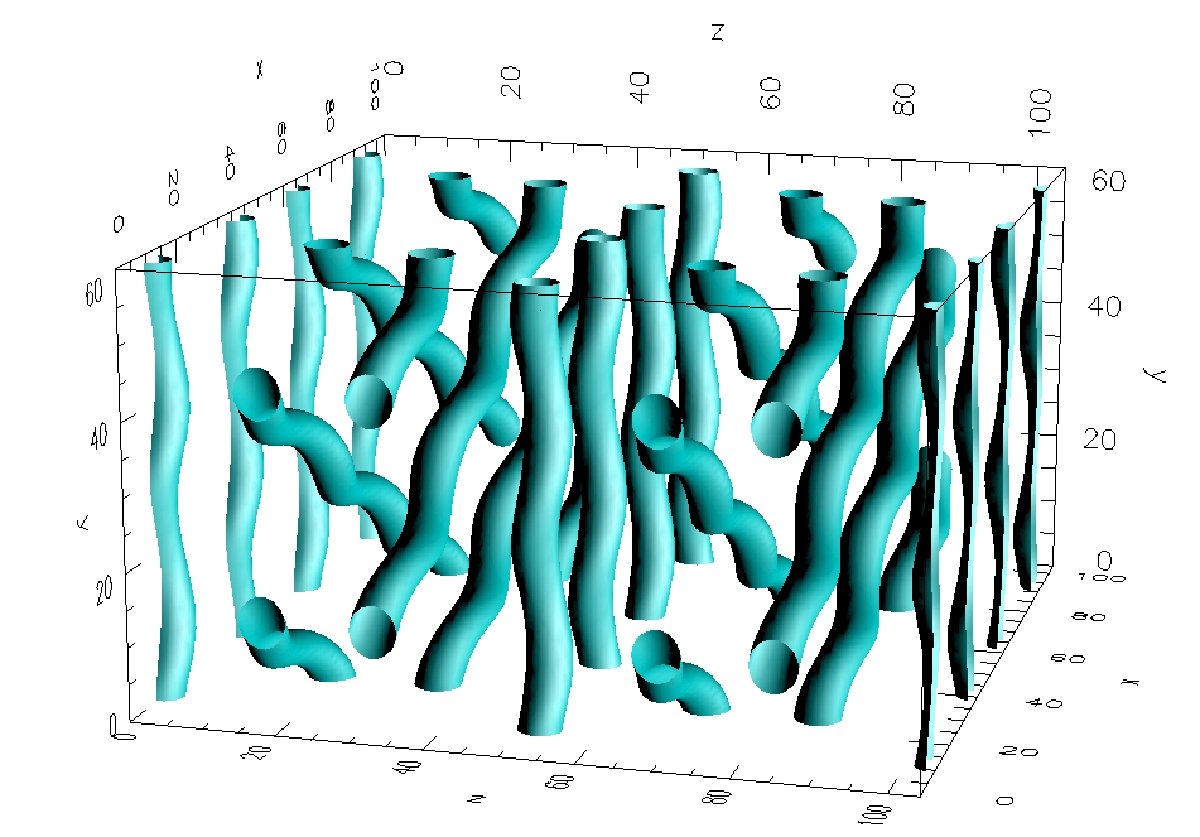}
\caption{Hexagonal-3Db BP in equilibrium state at $\tau=0.6, \kappa=1.5, e=0.3$.}
\label{fig12}
\end{figure}

\begin{figure}[h]
\includegraphics[width=0.5\textwidth]{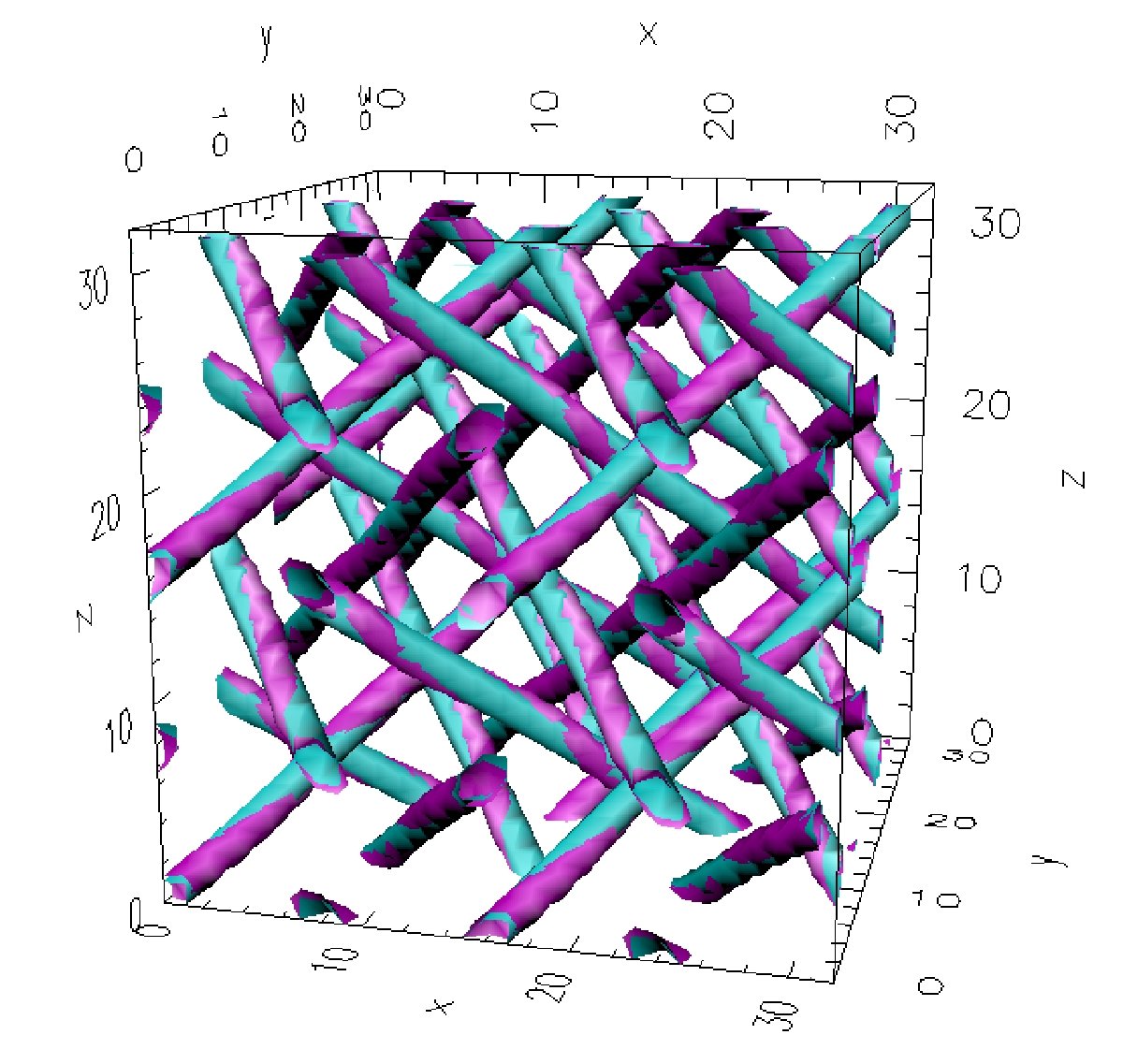}
\includegraphics[width=0.5\textwidth]{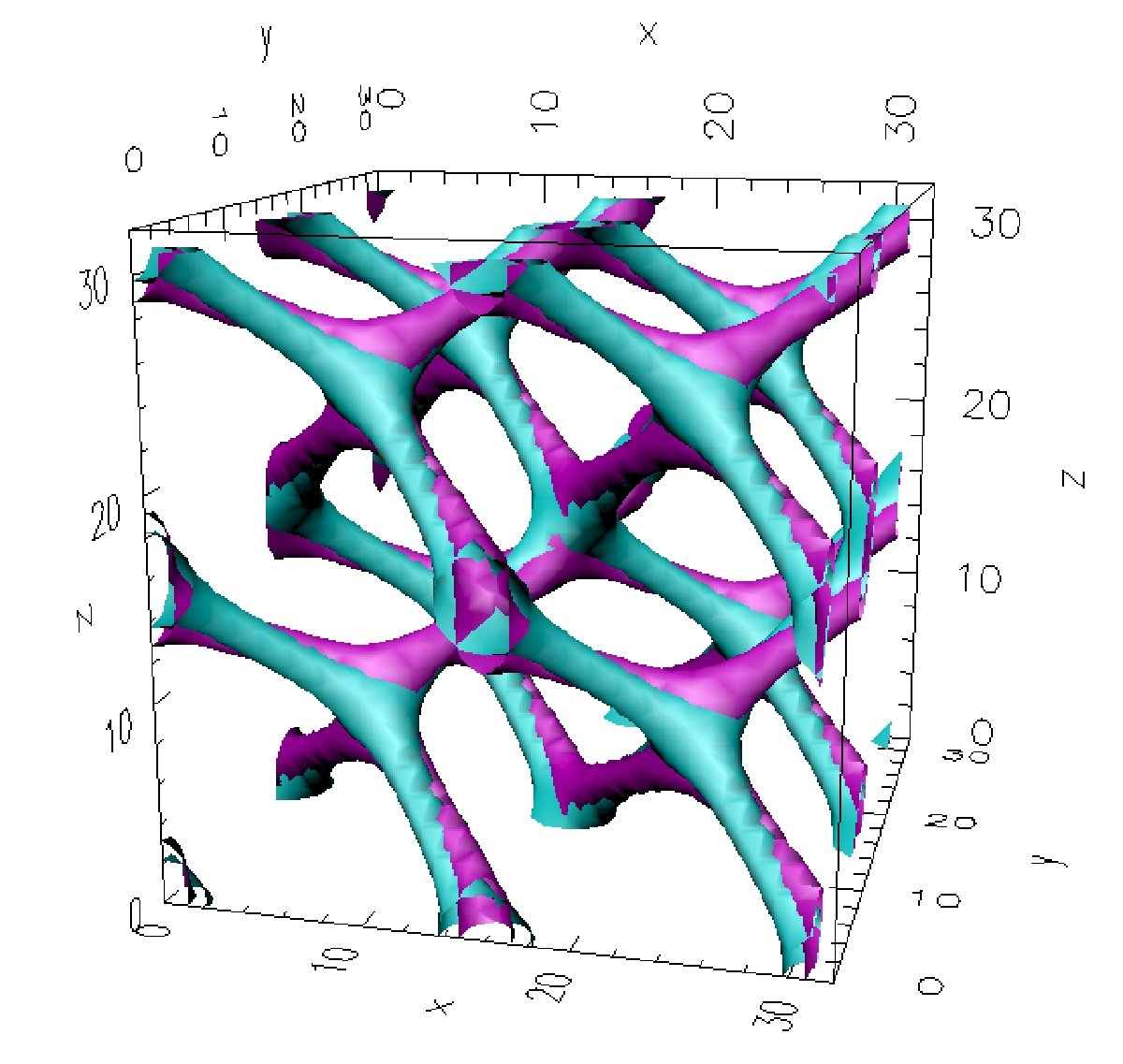}
\caption{BPI (top, at $\tau=-0.8, \kappa=1.5$) and BPII (bottom, at $\tau=0, \kappa=1.5$) at $e=0$ (cyan) and $e=0.3$ (magenta).}
\label{fig13}
\end{figure}

\end{document}